\documentclass[useAMS, usenatbib]{mn2e}

\usepackage{graphicx}	
\usepackage{aas_macros}
\usepackage{amsmath, amssymb}
\usepackage{array}
\usepackage[usenames,dvipsnames]{color}
\usepackage{multirow}
\usepackage{threeparttable}
\usepackage{comment}
\usepackage{float}

\newcommand{\myemail}{Mohammad.Nawaz@anu.edu.au}

\title[Jet-ICM interaction of Hydra A]{Jet-Intracluster Medium interaction in Hydra A. I Estimates of jet velocity from inner knots}
\author[M. A. Nawaz et al.]{
M.~A. Nawaz$^{1}$,\thanks{E-mail Mohammad.Nawaz@anu.edu.au} 
A.~Y. Wagner$^{2}$,
G.~V. Bicknell$^{1}$,
R.~S. Sutherland$^{1}$ 
and B.~R. McNamara$^{3}$
\\
$^{1}$ Research School of Astronomy and Astrophysics, The Australian National University, ACT 2611, Australia; \myemail \\
$^{2}$ Center for Computational Science, Tsukuba University, 1-1-1 Tennodai, Tsukuba, Ibaraki, 305-8577 Japan \\
$^{3}$ Department of Physics and Astronomy, University of Waterloo, Waterloo, ON, Canada}

\begin{document}

\date{Accepted 2014 July. Received 2014 May}

\pagerange{\pageref{firstpage}--\pageref{lastpage}} \pubyear{2014}

\maketitle

\label{firstpage}

\begin{abstract}
We present the first stage of an investigation of the interactions of the jets in the radio galaxy Hydra A with the intracluster medium. We consider the jet kinetic power, the galaxy and cluster atmosphere, and the inner structure of the radio source. Analysing radio observations  of the inner lobes of Hydra~A by \citet{taylor90} we confirm the jet power estimates $\sim 10^{45} \rm \> ergs \> s^{-1}$ derived by \citet{wise07} from dynamical analysis of the X-ray cavities.
With this result and a model for the galaxy halo, we explore the jet-intracluster medium interactions occurring on a scale of 10 kpc using
two-dimensional, axisymmetric, relativistic pure hydrodynamic simulations. A key feature is that we identify the three bright knots in the northern jet as biconical reconfinement shocks, which result when an over pressured jet starts to come into equilibrium with the galactic atmosphere. 
Through an extensive parameter space study
we deduce that the jet velocity is approximately $ 0.8 \,c$ at a distance $0.5 \ \rm kpc$ from the black hole. The combined constraints of jet power, the observed jet radius profile along the jet, and the estimated jet pressure and jet velocity imply a value of the jet density parameter $\chi \approx 13$ for the northern jet. We show that for a jet $\beta = 0.8$ and $\theta = 42^\circ$, an intrinsic asymmetry in the emissivity of the northern and southern jet is required for a consistent brightness ratio $\approx 7$ estimated from the 6cm VLA image of Hydra A. 
\end{abstract}


\section{Introduction}

Comprehensive radio and X-ray observations (see the reviews by \citealt{mcnamara07, mcnamara12}, and \citealt{fabian12}, and references therein) and numerical models \citep{gaspari11, dubois2010a} have established that the interactions between radio jets and the intracluster medium (ICM) counteract the cooling by X-rays in galaxy clusters, in which ``cooling flows'' would develop without the energy input by the AGN of the central cluster galaxy. This form of feedback, termed ``radio-mode'' feedback, is invoked in semi-analytic models and cosmological hydrodynamic simulations of galaxy formation to regulate the growth of the most massive galaxies and explain their deficit in present-day galaxy-luminosity functions \citep{croton06, okamoto2008b, dubois2013a}. 

The Hydra A cluster (Abell 780) is a well-studied, relatively nearby cool core cluster at a distance to the central radio galaxy of approximately 230 Mpc ($z$ = 0.054). There exists a wealth of radio and X-ray observations of the jets of Hydra A and of the ambient ICM \citep{taylor90, mcnamara00, david01}. Therefore, detailed models of the evolution of the radio jets in the Hydra A environment have the potential to provide valuable insights into the physics of radio-mode feedback.

Using high-resolution \textit{Chandra} data \citet{david01} showed that Hydra A is a cooling flow galaxy cluster with a mass accretion rate of approximately $\dot{M}$$\sim300 M_{\odot}\,\mathrm{yr}^{-1}$ beyond 30~kpc from the centre of the X-ray source. However, inside 30~kpc the mass accretion rate indicated by the X-ray spectroscopy drops sharply indicating that a heating mechanism is active near the cluster centre.

A discontinuity in the X-ray surface brightness and temperature profiles indicates the existence of a large scale shock front at $\sim200$ -- $300$ kpc \citep{nulsen05}. X-ray surface brightness deficiencies in the atmosphere were identified as a chain of X-ray cavities associated with radio bubbles \citep{wise07}.

\begin{figure*}
\centering
\includegraphics[width=\linewidth]{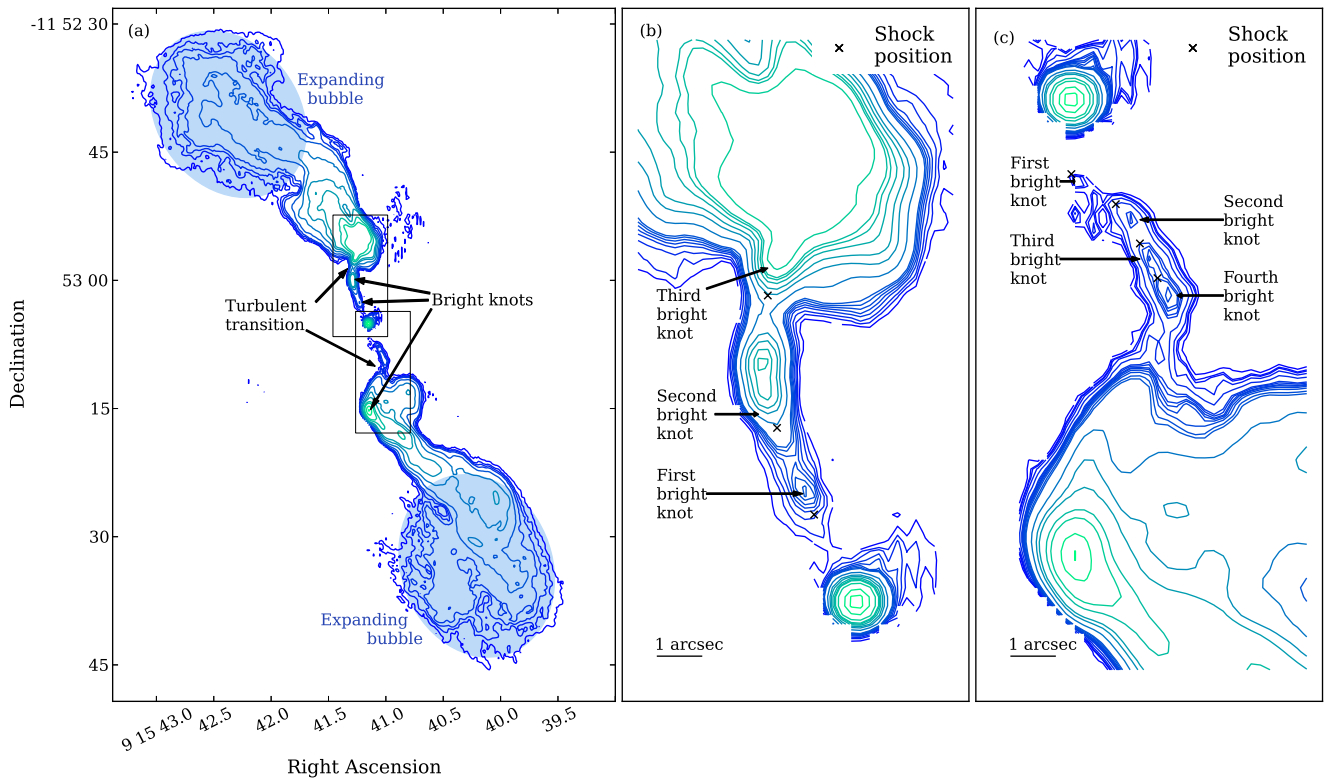}
\caption{\textbf{(a)} Radio intensity map of Hydra~A at 4.635 GHz. This figure is almost identical to Fig. 1 and Fig. 3 in \citet{taylor90}. Contour levels are at 1.5, 2.7, 3.7, 5.1, 10, 21, 37, 51, 103, 154, 311, and 466 mJy arcsec$^{-2}$. 
The elliptical areas outline the approximate volume of the corresponding X-ray A and B cavities and are used to estimate the contribution to the jet kinetic power. \textbf{(b)} Zoom-in of the top rectangular region in a) showing the bright knots in the northern jet. Contours are at 1.5, 2.7, 3.7, 5.1, 6.3, 7.5, 8.8, 10, 21, 37, 51, 72, 90, 103, 154, 311, and 466 mJy arcsec$^{-2}$. \textbf{(c)} Zoom-in of the bottom rectangular region in a) showing the bright knots in the southern jet. Contours are at 1.5, 2.0, 2.2, 2.7, 3.7, 5.1, 5.5, 6.0, 6.3, 6.8, 7.5, 8.8, 10, 21, 37, 51,  72, 90, 103, 154, 311, and 466 mJy arcsec$^{-2}$.
(A colour version of this figure is available in the online journal)}
\label{fig1}
\end{figure*}

Hydra A has also been observed at a wide range of radio frequencies. Low frequency Very Large Array (VLA) observations reveal the remnants of the early epochs of radio activity \citep{lane04}, while GHz observations reveal active jets and inner radio lobes in the central $\sim50$ kpc \citep{taylor90}. Fig.~\ref{fig1}a is a reproduction of the 4.635 GHz image in Fig.~1 of \citet{taylor90}. In the inner region of the radio source both jets flare, producing plumes at a deprojected distance of approximately 10 kpc from the core, assuming an inclination angle $\theta=42^\circ$ derived from rotation measure asymmetries \citep{taylor90,taylor93}.

In the northern jet, a bright knot at a deprojected distance of $\sim 7$ kpc from the core is apparent just before the jet flares. At approximately 3.7~kpc from the core, another fainter knot is visible. A more bright third knot within the turbulent region at approximately 11.7~kpc is also visible. In the southern jet four bright knots are apparent at $\sim2.50\, \ 3.90\, \ 5.40\, \rm \ and \ 6.70 \ kpc$ from the core. The bright knots and the flaring points are enlarged and clearly seen in the zoomed-in region shown in panels b and c of Fig.~\ref{fig1}. The trajectories of the northern and southern jets in the inner 10 kpc from radio core exhibit an S-shaped structure, which persists in morphology of the plumes. The rotationally symmetric S-structure is also evident in the low frequency images at 74 MHz and 330 MHz \citep{lane04}.

The spatial anti-correlation between radio and X-ray emission in Hydra A strongly indicates that the radio jets impact large volumes of the ICM gas and regulate the cooling flow in Hydra A. The correlation between jet power and X-ray luminosity in the \citet{birzan04} sample support such a scenario in cooling flow clusters in general. 

In recent years, several models of the Hydra A radio source and the ICM have been presented. \citet{simionescu09} proposed, using hydrodynamic simulations, that the interaction of very powerful jets ($\sim6\times10^{46}$ erg s$^{-1}$) with a spherically symmetric hydrodynamic environment can reproduce the observed large scale shock front with a  Mach number $M \sim 1.3$. In order to explain an offset of 70 kpc between the centre of the shock ellipse and the cluster core, the interaction was deemed to take place in two stages: First, active jets propagate through a hydrostatic environment within 100 kpc from the core; Second, the jets turn off and buoyant bubbles rise through a background environment that has a bulk velocity of 670 km s$^{-1}$ relative to the central galaxy. In that study, the base of the jet in the hydrodynamic simulations is located at approximately 10 kpc from the core where the jet radius is approximately 6 kpc. The inner 10 kpc region, where the jet has not yet transitioned to a turbulent flow, was not explored. 

\citet{rafaelovich12} also modelled Hydra A using axisymmetric, hydrodynamic simulations and showed that a single outburst can produce a series of X-ray deficient bubbles. In their model, the vortex shedding and the Kelvin-Helmholtz instabilities at the contact discontinuity of the shocked ICM and the shocked jet plasma are responsible for multiple X-ray cavities.

So far, the theoretical modelling of Hydra A has focused on the large-scale structures, such as the cavities and the shock fronts bounding the expanding bubbles. However, no numerical simulations have related the outer structure of the radio source to the structure within $\sim10$ kpc of the radio core. One feature in particular that demands attention is the jet-plume transitions in the northern and southern jets, which mark a dramatic change in the flow properties of the jets.

In this series of papers we address several aspects of the radio and X-ray properties of Hydra A. These include the energy budget of the radio source and its inner 10 kpc structure. In this first paper we concentrate on the physics of the jets in the first 10 kpc and develop the following model: Initially the supersonic jets are conically expanding and over-pressured with respect to the interstellar medium (ISM) as they emerge from the core. As their pressure decreases, they start to come into pressure equilibrium with the (ISM) and produce a series of biconical shocks, which are visible as knots. In conjunction with this process the jet boundary oscillates. The spacing of the knots and the variation in jet radius are sensitive to the overpressure ratio and the velocity of the jet (for a given jet kinetic power) enabling us to estimate both. In addition, the jets decelerate at each shock and make a transition to turbulence  \citep{bicknell84a} consistent with the formation of plumes in the radio image (see \ref{fig1}). We note however, that this transition is only indicative since turbulence is a three-dimensional phenomenon, which is not fully probed by the axisymmetric simulations that we employ here. Paper~II considers three-dimensional phenomena.

As noted, in paper II, we utilise three-dimensional simulations and model the dynamical interaction of the jets and the ICM utilising a precessing jet model. In a third paper we address the detailed radio and X-ray emission features predicted by our model.

The paper is structured as follows.
In \S~\ref{jet_kinetic_power} we discuss estimates of the jet kinetic power and composition. The description of our model of the interaction of initially conical and ballistic Hydra A jets with the cluster environment is presented in \S~\ref{s:model}. In \S~\ref{s:cluster} we construct an analytic model of the
density, temperature and pressure of the Hydra~A ICM from fits to published X-ray data. The parameters derived in these two sections are used in the numerical simulations, which we present in \S~\ref{s:code} and \S~\ref{s:sims}. We summarise and discuss our results in \S~\ref{s:discussion}.

%
%
\section{Estimates of Jet kinetic power} \label{jet_kinetic_power}

In order to construct physically realistic models of the interaction of the radio-jets with the environment of Hydra~A, we require good estimates of the jet kinetic power and the spatial profiles of density, temperature and pressure in the cluster atmosphere. Previous estimates of the jet power \citep{nulsen05,wise07} are based on X-ray observations of the outer shock and the cavities produced by the radio source. In this section we both supplement and confirm these estimates by utilising radio data of the inner lobes of Hydra~A. 

%
%
\subsection{Jet power based on a model for the outer shock}\label{s:powsh}
\citet{nulsen05} focused on the outer shock evident in the X-ray image and used a spherically symmetric hydrodynamic model of a point explosion in an initially isothermal and hydrostatic environment to produce theoretical X-ray surface brightness profiles for Hydra~A. Their best fit to the observed X-ray brightness profile gives a shock age $\sim 1.4 \times 10^8 \, \rm Myr$ and an explosion energy $\sim 10^{61} \, \rm erg$. The estimated power of the outburst is $\sim 2 \times 10^{45} \, \rm erg \, s^{-1}$. On the basis of this model, we would associate $10^{45} \> \rm erg \> s^{-1}$ with each jet.

%
%
\subsection{Jet power based on X-ray cavities}\label{s:powx}

\citet{wise07} used the observations of three pairs of X-ray cavities revealed in \textit{Chandra} images to estimate the power of the Hydra~A jets. The inner cavities A and B correspond to the 4.6~GHz radio lobes \citep{mcnamara00}, the cavities C and D correspond to the middle lobe in the 1.4 GHz radio image \citep{lane04} and the outer cavities E and F correspond to the outer lobes in the 330 MHz image \citep{wise07}. Wise et al. consider that the three cavities on each side are interconnected and use the sum of the enthalpies, $h_\mathrm{tot}=\gamma p_{\rm lobe}V_{\rm lobe}/(\gamma -1)$, where $\gamma$ is the polytropic index, $p_{\rm lobe}$ is the pressure of the lobe and $V$ is the volume of the cavity, in all cavities to calculate a total outburst energy. From this, the combined jet power is calculated, $P_\mathrm{jet}=4 p_{\rm lobe} V/t_{\rm cav}$, where $t_{\rm cav}$ is the age of each cavity. The average of three different cavity age estimates was used: the time required for the X-ray cavity to reach its present position if moving at the sound speed, the refilling time of the X-ray cavity, and the time required for the X-ray cavity to rise buoyantly to the present position. Assuming pressure equilibrium of the lobes with the atmosphere they obtained powers for the inner and middle lobes of $\sim 2 \times 10^{44} \> \rm erg \> s^{-1}$ and for the outer lobes, $\sim 6 \times 10^{44} \> \rm erg \> s^{-1}$, which gives a combined jet power $\sim 2 \times 10^{45} \> \rm erg \> s^{-1}$. 

The authors find that a power of $1\times10^{45}$ erg s $^{-1}$ for the northern jet is consistent with the supposition that the jet is still filling the outermost of the X-ray cavities at $\sim200$ kpc (the corresponding radio lobe is visible at 330 Mhz) and driving the large-scale shock. An independent estimate of the jet power from the expansion rate of the outermost cavity, assuming a self-similar evolution of the radius of the cavity wall and the large scale shock agrees with their first estimate to within a factor of 2. This value of the jet power $1\times10^{45}$ erg s $^{-1}$ is also consistent with the estimate of the jet power obtained by \citet{nulsen05} noted above. 

\begin{table}
\caption{Parameters for the determination of the lobe minimum energy.} 
\centering
\begin{tabular}{l c c}
 \hline  \hline
 Parameter & Value \\
 \hline 
 Electron spectral index, $a$  &  2.4 \\
 Lorentz factor lower cutoff, $\gamma_1$   &  100                       \\ 
 Lorentz factor upper cutoff, $\gamma_2$ &   10$^6$ \\  
 Central surface brightness, $I_{\nu}$       & 10 mJy arcsec$^{-2}$	 \\ 
 Plasma depth, $L$    & 20 kpc (northern lobe) \\ 
                          & 22 kpc (southern lobe) \\ \hline
\end{tabular}
\label{radio_data}
\end{table}


%
%
\subsection{Estimates of the jet power from synchrotron minimum energy}\label{s:powr}

We revisit the calculation of the cavity powers of the two innermost cavities \citep[see][]{wise07} by using the synchrotron minimum energy estimate for the pressure and synchrotron ages of the lobes. The main difference between this method and that using the X-ray cavities is that the former introduces a strong dependence of the lobe pressure on the particle content of the lobe, whereas the X-ray cavity pressure only depends weakly on the particle content through the adiabatic index.

The work by \citet{croston05a} on the lobes of classical double (FRII) radio galaxies shows that using a synchrotron minimum energy estimate is a feasible approach. However, since Hydra~A is an FRI source this requires further justification. \citet{croston05a} used observations of the inverse Compton emission in their sample to show that the lobes are close to equipartition when the inverse ratio of energy in relativistic electrons/positrons to that of ``other'' particles, $k=0$. They use this fact to rule out an energetic relativistic proton component since the existence of such a component would imply that the magnetic field is in equipartition with the relativistic electrons only. While the authors did not state this directly, their argument can also be used to exclude an energetic thermal component. This is evident for powerful FRII sources, for which we do not expect much entrainment to occur. However, we argue that the turbulent processes leading to equipartition are independent of the plasma composition and that in the case where the plasma has a substantial thermal content these processes also lead to a minimum energy state between \emph{all} particles and the magnetic field. This conclusion is supported by the work of \citet{birzan08} discussed below.     

We have approximated the shapes of the lobes with ellipsoidal volumes as shown by the shaded elliptical regions in Fig.~\ref{fig1}a; the plasma depth is taken to be equal to the minor axis, $L$. The lobe centres are located at $\sim 30$~kpc from the core. Let $I_\nu$ be the central surface brightness of each lobe, where $\nu=4.6$ GHz is the frequency of the \citet{taylor90} observations. Let $m_{\rm{e}}$ be the electron mass, $e$ the elementary charge, $a$ the electron index, $\alpha=(a-1)/2$ the spectral index, and $\gamma_1$ and $\gamma_2$ the lower and upper cutoff Lorentz factors, respectively. Then, the minimum energy magnetic field (in Gauss) \citep[e.g][]{bicknell13a} of the synchrotron radiating plasma is given by: 

\begin{equation}
B_{\rm{min,E}} = \frac{m_{\rm{e}} c}{e}\left[ \frac{a+1}{2}(1+k)C^{-1}(a)\frac{c}{m_{\rm{e}}}f(a,\gamma_1, \gamma_2)\frac{I_{\nu}\nu^{\alpha}}{L} \right]^{\frac{2}{a+5}}\;,
\label{e:bmin}
\end{equation}
where 
\begin{equation}
f(a, \gamma_1, \gamma_2) = (a-2)^{-1} \gamma_1^{-(a-2)} \left [ 1 - \left( \frac{\gamma_2}{\gamma_1}\right)^{-(a-2)}\right]\:,
\end{equation}
and 
\begin{eqnarray}
C(a) &=& 3^{a/2}2^{-(a+7)/2}\pi^{-(a+3)/2} \nonumber  \\
&& \times  \frac{\Gamma\left(\frac{a}{4}+\frac{19}{12}\right)\Gamma\left(\frac{a}{4}-\frac{1}{12}\right)}{a+1} \> 
\frac{\sqrt{\pi}}{2}\frac{\Gamma\left(\frac{5+a}{4}\right)}{\Gamma \left(\frac{7+a}{4}\right)}\;.\label{eq:ca}
\end{eqnarray}
In Eqn.~\eqref{eq:ca} $\Gamma$ is the Gamma-function. Values adopted for $a$, $\gamma_1$, $\gamma_2$, $I_{\nu}$, and $L$ are shown in Table \ref{radio_data}. We choose a spectral index $\alpha\approx0.7$ (hence $a=2.4$), which is representative of the low frequency spectral index of the radio emission \citep{cotton09}.
We choose a lower Lorentz cutoff $\gamma_1=100$, in view of numerous studies of radio galaxies finding $\gamma\sim100-10^3$  \citep{carilli91,  hardcastle01, godfrey09}. Also, $\gamma_2\approx10^6$, since this corresponds to emission frequencies well above the microwave range. Minimum energy estimates are insensitive to $\gamma_2$ and only weakly dependent on $\gamma_1$.

The minimum total (particles + field) energy density of the lobe is given by 
\begin{equation}
\varepsilon_{\rm{tot}} = \varepsilon_{\rm p} + \frac {B_{\rm min,E }^2}{8 \pi} = \frac{a+5}{a+1}\frac{B_{\rm{min,E}}^2}{8 \pi}\;,
\end{equation}
and the total pressure of the lobe corresponding to the minimum total energy density is
\begin{equation}
p_{\rm{tot}}=\frac{1}{3}\varepsilon_\text{e}(1+k)+\frac{B_\text{min,E}^2}{8 \pi}\:.
\label{total_pressure}
\end{equation}
The major uncertainty in this calculation arises from the lack of observational constraints on the parameter $k$. The value of $k$ determines whether the lobe (in equipartition) is over-pressured or under-pressured with respect to the environment, and later in this section we discuss the range of values for $k$ applicable to the Hydra~A radio lobes. Estimates of $B_{\rm{min,E}}$ and $p_{\rm{tot}}$ are given in Table \ref{jet_parameters} for three values of $k$. 

We estimate the age of the source from the curvature in the spectrum derived by \citet{cotton09} who showed that the spectra of the inner lobes steepen for frequencies $\ga 300 \> \rm MHz$. Let $B=B_\mathrm{min,E}$ be the minimum energy magnetic field and $\nu_{\rm{b}} \approx 300 \> \rm MHz$ be the break frequency, then the synchrotron age of the source is 
\begin{equation}
t_{\rm{rad}} \approx \frac{3^{5/2}}{8 {\pi}^{1/2}} \left(\frac{m_e^3 c^5}{e^3}\right)^{1/2} B^{-3/2}\nu_{\rm{b}}^{-1/2}\:.
\end{equation}
Hence, the power associated with each of the inner cavities is 
\begin{equation}
P_{\rm cav} = \frac{\gamma}{(\gamma -1)} \frac{p_{\rm lobe} V_{\rm lobe}}{t_{\rm rad}}.
\label{p_cav}
\end{equation}
We use the total pressure $p_{\rm{tot}}$ for minimum energy conditions as the lobe pressure. Values of $P_{\rm{cav}}$ for different values of $k$ are given in Table \ref{jet_parameters}.

%
%

\begin{table*}
\caption{Parameters calculated for three values of $k$ using synchrotron minimum energy for both lobes of Hydra A. }
\label{minimum_energy}
\centering
\begin{threeparttable}
\begin{tabular}{*{7}{c}}
\hline \hline
   $k$ & $B_{\rm{min,E}}$  & $\varepsilon_{\rm{tot}}$  & $p_{\rm{tot}}$ &  $p_{\rm{tot}}/p_{\rm{a}}$ & $P_{\rm{cav}}$ & $t_{\rm{rad}}$    \\
       & (10$^{-6}$ Gauss)  & ($10^{-10}$ erg cm$^{-3}$) &  ($10^{-10}$ dyne cm${^{-1}}$) & & ($10^{44}$ erg s$^{-1}$) & (Myr)  \\ 
             \multicolumn{7}{c}{Northern Lobe} \\ \hline
  0 & 22 & 0.4 & 0.3  & 0.2 & 0.1 & 49  \\ 
  10 & 42 & 1.5 & 1.2 & 0.9 &  1.8 &  18 \\ 
  100 & 76 & 4.9 &  3.9 & 3.0 & 14.5  & 7  \\
  \hline
   \multicolumn{7}{c}{Southern Lobe} \\ \hline
       0 &  21 & 0.4 & 0.3 & 0.2 &  0.1  & 51 \\
       10 & 40 & 1.4 & 1.1 &  0.9 &  2.0 & 19 \\ 
       100 & 73 & 4.6 & 3.7 & 3.0 &  16.3   & 8 \\
\hline
\end{tabular}
\label{jet_parameters}
\end{threeparttable}
\end{table*}

Table \ref{minimum_energy} shows, for both the northern and southern lobes, the estimation of the minimum energy magnetic field, $B_{\rm{min,E}}$, the total energy density $\varepsilon_{\rm tot}$, the total pressure of the lobe, $p_{\rm{tot}}$, the ratio between the total lobe pressure and the atmospheric pressure $p_{\rm{tot}}/p_{\rm{a}}$, the cavity power for $\gamma = 4/3$, and the radiative ages of the lobes for values of the parameter $k=0, \ 10 \rm \ and \ 100$.

For the same value of $k$, the cavity powers of the northern and southern lobes are comparable. Moreover, for $k=10$ the lobes are in approximate pressure equilibrium with the atmosphere and the cavity powers ($1.8 \times 10^{44}$ and $2.0\times 10^{44} \> \rm ergs \> s^{-1}$ respectively) agree with the \citet{wise07} estimates of $2.1\times 10^{44} \> \rm ergs \> s^{-1}$ and $2.0\times 10^{44} \> \rm ergs \> s^{-1}$ respectively. For $k=0$ the lobes appear to be significantly under-pressured and for $k=100$ significantly over-pressured.

This high value of $k$ is consistent with the estimate of $k=13$ by \cite{birzan08}. They have also estimated this parameter for the inner lobes of Hydra A by invoking pressure equilibrium with the atmosphere. 

The major uncertainty associated with these radio-based estimates of the cavity power is that there is no direct estimate of the lobe pressure and we have assumed that the lobe pressure is determined by the total pressure of the lobe when the lobe is in its minimum energy state. This assumption gives a lower limit of the lobe energy, and hence a lower limit on the cavity power.

The estimation of the power associated with the inner radio lobes given in \S~\ref{s:powr} and the power of the corresponding X-ray cavities given in \S~\ref{s:powx} for a nearly pressure equilibrium situation are consistent with the cavity power estimates for the northern and southern lobes respectively. This indicates that the result for the total jet power obtained by \citet{wise07} is reliable and provides a sound basis for numerical models of Hydra A. We therefore adopt a jet power of $10^{45} \> \rm erg \> s^{-1}$ as our value in the simulations presented in \S~\ref{s:sims}.

%
%
\section{Jet parameters} \label{s:model}

\begin{figure}
\includegraphics[width=\linewidth]{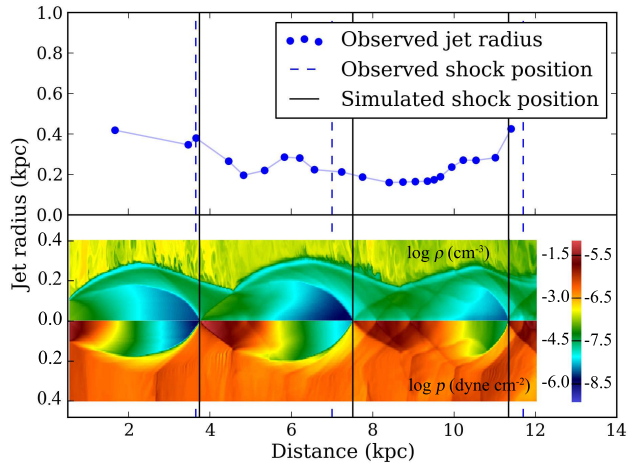}
\caption{Plot of the jet radius of the northern jet and the location of shocks as a function of the deprojected distance from the core. The radius is estimated from the deconvolved FWHM (see text). The vertical dashed lines represent the location of the southern edge of the first two bright knots in the northern jet, which correspond to the assumed locations of the shocks. In the bottom two panels the simulated logarithmic density and pressure slices show the periodically expanding and reconfining morphology and the shocks produced in our best-fit model for the northern jet. Both the radius plot and the images are stretched in the radial direction, emphasising the wave-like nature of the jet boundary. The vertical solid lines represent the shock positions in the simulations. The colourbar represents $\log \rho$ on the left and $\log p$ on the right.}
\label{f:radius}
\end{figure}

In this section we describe the selection of the initial jet parameters, the jet cross-sectional area $A_{\rm jet} = (\pi r_{\rm jet}^2$, where $r_{\rm jet}$ is the jet inlet radius), the jet pressure $p_{\rm jet}$, the jet density parameter $\chi = \rho_{\rm jet} c^2 / (\epsilon_{\rm jet} + p_{\rm jet})$, where $\rho_{\rm jet}$ and $\epsilon_{\rm jet}$ are the rest mass density and the energy density of the jet respectively and the jet Lorentz factor $\Gamma = (1- \beta^2)^{-1/2}$.  These parameters are assigned so as to be consistent with the expression for the jet power:
\begin{equation}
P_{\rm{jet}} = \frac{\gamma}{\gamma -1} c p_{\rm{jet}}\Gamma^2\beta A_\mathrm{jet}\left(1+\frac{\Gamma -1}{\Gamma}\chi\right)\:.
\label{jet_power}
\end{equation} 
\citep{sutherland07a}

\subsection{Magnetic field}

In our simulations we neglect the magnetic field. Is this a reasonable approximation given the popular notion that jets may be collimated by the toroidal field, which develops as a result of the rotation of the flow ejected from the accretion disk \citep{blandford82a} or from the ergosphere \citep{blandford77a}? In this case one expects the magnetic and particle pressures to be comparable. Moreover, this would argue against the assumption of invoking an over-pressured jet on the parsec-scale (see below). Self-collimation by a toroidal magnetic field is an appealing mechanism for the region of jets just outside the Alfven surface. However, the fact that the jet expands by a factor of over 200 between the parsec scale and the kiloparsec scale indicates that self-collimation does not occur in this region. For example the self-similar models of \citet{li92a} and \citet{vlahakis03a} indicate that asymptotically the flow becomes cylindrical when the jet is magnetically collimated. A different model has been proposed by \citet{spruit11a}, who has argued that three-dimensional effects lead to reconnection of the magnetic field and that the loss of magnetic energy produces a pressure gradient, which is responsible for the acceleration of jets to high Lorentz factors. \citet{moll09a,moll10a} has carried out numerical simulations based on this concept, in the context of protostellar jets. There is also observational support for 
sub-equipartition magnetic fields on the sub-parsec scale in a substantial fraction of gamma ray blazars. In a recent paper \citet{zhang14a} modelled the spectral energy distributions of a number of BL~Lac objects and flat spectrum radio quasars (FSRQs) and found that they divide along the line magnetic~power = electron~power with most of the BL~Lac objects being below this dividing line (see their Fig. 13(b)). The respective powers are proportional to the energy densities of the various components (their section 4) so that the ratio of the magnetic power to electron power informs us of the ratio of the respective energy densities. Hence, the magnetic energy densities in many of the BL~Lac objects are well below the electron energy density (but with some members of the sample approaching equality). Thus there is good justification, in the first instance, for neglecting the magnetic field with the implication that the beamed counterpart of Hydra~A would be a BL~Lac object rather than a quasar.

What values of the jet density parameter, $\chi$ are relevant in this context? Two main options for jet composition are generally discussed -- electron-positron or electron-proton. Let $m_e$ be the electron mass and $m_+$ the mass of the positively charged component, $m_e$ for a positron and $m_p$ for a proton. The parameter $\chi$ is then given by: 
\begin{equation}
\chi = 0.75(a-2)(a-1)^{-1} \, \frac {m_+}{m_e} \gamma_1^{-1}
\end{equation}
where $a$ and $\gamma_1$ are defined in \S~\ref{s:powr}. Note that, for an electron positron jet with $a=2.4$ and $\gamma_1 \ga 10$, $\chi \ll 1$. The theory of jet production from black holes \citep{blandford77a} and X-ray observations of the lobes of both FR1 and FR2 radio galaxies \citep{croston05a,croston14a} make the concept of electron-positron jets appealing. However, the issue of jet composition is by no means settled. In an electron-proton jet, low values of $\chi$ require the low energy cutoff, $\gamma_1 \gg 1$.

\subsection{Over-pressured jets}

A key feature of our jet model is that the bright knots beginning at $\sim3.7$ and 7.0 and 11.7 kpc from the core in the northern jet and at $\sim 2.5$, 3.9, 5.4 and 6.7 kpc in the southern jet are the result of  consecutive biconical shocks following recollimation of over-pressured jets. We have identified the points where the surface brightness gradient markedly increases, as the location of the upstream side of each knot. (See Fig.~\ref{fig1}.) The third knot in the northern jet occurs just as the jet merges into the lobe so that we might expect the location of this knot to be affected somewhat by the jet's transition to turbulence.  

\citet{norman82} first drew attention to the production of biconical and normal shocks (Mach discs) in over-pressured astrophysical jets. An initially over-pressured jet expands laterally and its thermal pressure and ram pressure decreases with distance along the direction of propagation. When the jet pressure reaches the ambient pressure the jet begins to recollimate. The jet periodically expands and recollimates, producing a series of biconical or normal shocks along the jet axis. This phenomenon had been known to laboratory hydrodynamicists for some time and \citet{birkhoff57a} associated it with the \emph{natural wavelength} of a supersonic jet $\Lambda$. 

It is feasible that the Hydra A jets are initially over-pressured since the minimum energy pressure in the pc-scale northern jet, 27 pc from the central black hole \citep{taylor96} is $1.33\times10^{-7}$ and $1.26\times10^{-7} \rm \ dynes \ cm^{-2}$ for $\beta=0.2$ and 0.9 respectively. (A jet diameter of 26 pc was used in these estimates). These minimum energy pressure estimates are about a factor of 200 times higher than the central pressure $\approx 6.6\times10^{-10} \rm \ dynes \ cm^{-2}$ of the modelled interstellar medium (see \S~\ref{s:cluster}). Moreover, using these pressures underestimates the jet kinetic power at $ \approx 2.4 \times10^{44} \ \rm erg \ s^{-1}$ for $\beta =0.8$ and $\chi \sim 10^{-2}$ compared to the value $10^{45} \ \rm erg \ s^{-1}$ used in our models by a factor $\approx 4$. Therefore the value of the jet kinetic power of our models implies a jet pressure $6~\times p_{\rm min} \approx 7.0 \times 10^{-7} \rm \ dynes \ cm^{-2} $ at 27~pc. This is approximately 100 times the central atmosphere pressure. If we assume that the jet expands adiabatically, i.e., the jet pressure decreases with the jet radius according to $p_{\rm jet}\propto r_{\rm jet}^{-8/3}$, we obtain an over pressured jet $5~p_{\rm a}$ at 0.5~kpc  from the core (where we initialise our jet in the computational domain) with a jet radius 100 pc.

Interpreting the jet as over-pressured on the parsec scale implies that from the parsec to the kiloparsec scale it is freely expanding. We also note here that, in a detailed analysis of protostellar jets \citet{cabrit07a} has concluded that those jets are initially magnetically collimated but are freely expanding at some distance ($\sim 50 \> \rm AU$) from the star. Of course, these scales are not directly commensurable with Hydra~A, but a long held view is that the physics of protostellar and AGN outflows are similar in many respects.

Our proposition of the jet bright knots as biconical shocks is further reinforced by the observed wave-like nature of the northern jet boundary. Fig~\ref{f:radius} shows the radius profile (dots) of the northern jet, which we obtain by assuming the jet as a homogeneous cylinder and utilising the deconvolved FWHM of the jet \citep{taylor90} $\Phi_{\rm jet}$ together with $r_{\rm jet} = \Phi_{\rm jet}/ \sqrt{3}$. In order to illustrate the association of biconical shocks with the sinusoidal radius profile we attach the logarithmic density and pressure images (panels marked with $\log\rho$ and $\log p$ respectively) of one of our best fit models Ciii for the northern jet. In the simulated radius profile we see the jet boundary oscillates and at $\sim 0.7 \ \rm kpc$ before each radius minimum biconical shocks appear. These are clearly indicated by the large increase in pressure. The observed and simulated shock locations are marked with dashed and solid vertical lines respectively.

We construct models of the northern jet for which data on the jet FWHM are more complete. Our modelling strategy for this jet is as follows. We conduct a parameter space study searching for numerical models which can successfully reproduce the correct shock locations and the radius profile of this jet. 

In our axisymmetric numerical models of the jet-ICM interaction we deal with straight jets whereas the Hydra A jets are curved. However since the curvature of the jets are modest within the central 10~kpc, we expect an approximation by a straight jet to be reasonable.

As stated above, in order to model the jets of Hydra A we require five jet parameters, the jet kinetic power $P_{\rm{jet}}$, the initial jet radius $r_{\rm{jet}}$, the initial jet pressure $p_{\rm{jet}}$, the initial jet velocity $\beta$ (in units of the speed of light), and the jet density parameter $\chi$, of which four are independent. In the previous section we established a value for the jet kinetic power $10^{45} \rm erg \ s^{-1}$. In the following we describe how we choose the other three independent jet parameters and set their values. 

Our first parameter is the jet kinetic power, which is reasonably well-determined by the radio and X-ray observations. The jet radius is our second parameter; this affects the downstream scale of the oscillating jet boundary and is not known \emph{ab initio}. The third parameter is the jet pressure ratio; this affects both the amplitude of the radial oscillations and the knot spacing. The fourth parameter is the jet velocity, $\beta$. Then the parameter $\chi$ is determined by solving equation~(\ref{jet_power}) for $\chi$, that is,
\begin{equation}
\chi = \frac{\Gamma}{\Gamma -1}\left( \frac{\gamma-1}{\gamma}\frac{P_{\rm{jet}}}{c p_{\rm{jet}} \Gamma^2\beta A_{\rm{jet}}} -1 \right)\:.
\label{chi2}
\end{equation}

Referring to the expression for the natural wavelength $\Lambda$ of a supersonic \emph{non-relativistic} jet in near pressure equilibrium:
 \begin{equation}
\Lambda/r_{\rm jet} \approx 2.6 \sqrt{M^2 - 1}
\label{e:birkhoff}
\end{equation}
\citep{birkhoff57a}, we note that our selection of the velocity and density parameters is equivalent to defining the Mach number $=(2+3 \chi)^{1/2} \Gamma \beta$ \citep{bicknell94a}. 

Following \citet{komissarov98} and \citet{krause12} we model the jet as ballistic and conically expanding in the first 
0.5~kpc, which represents the base of the computational domain. \citet{komissarov98} used an identical setup in their simulations to show that an initially conical jet may be collimated by the ambient pressure. \citet{krause12} performed simulations, also with identical initial conditions to provide a theoretical basis for the FRI/FRII classification of radio sources based on the half cone angle of the initial jet cone.

To summarize, we set up our simulations with an initially over-pressured (in one case equilibrium pressure) conically expanding jet with cross-sectional radius $r_{\rm{jet}}$ and centre at $(r, z) = (0, 0.5)$ kpc, where $r$, and $z$ are the radial and height coordinate of our axisymmetric cylindrical domain.  The independent jet parameters are jet power $P_{\rm{jet}}=10^{45}\,\rm erg \, s^{-1}$, jet radius $r_{\rm{jet}}$, inlet jet pressure $p_{\rm{jet}}$, inlet jet velocity $\beta$. The remaining jet parameter $\chi$ is determined from Eqn.~\ref{chi2}
The components of the jet velocity at a points $(r, z)$ within the initial conically expanding jet cross-section are $v_r = \beta z/\sqrt{r^2 + z^2}$ and $v_z =\beta r/\sqrt{r^2 + z^2}$.

%
%
\section{Cluster Environment} \label{s:cluster}

\begin{figure}
\includegraphics[width=\linewidth]{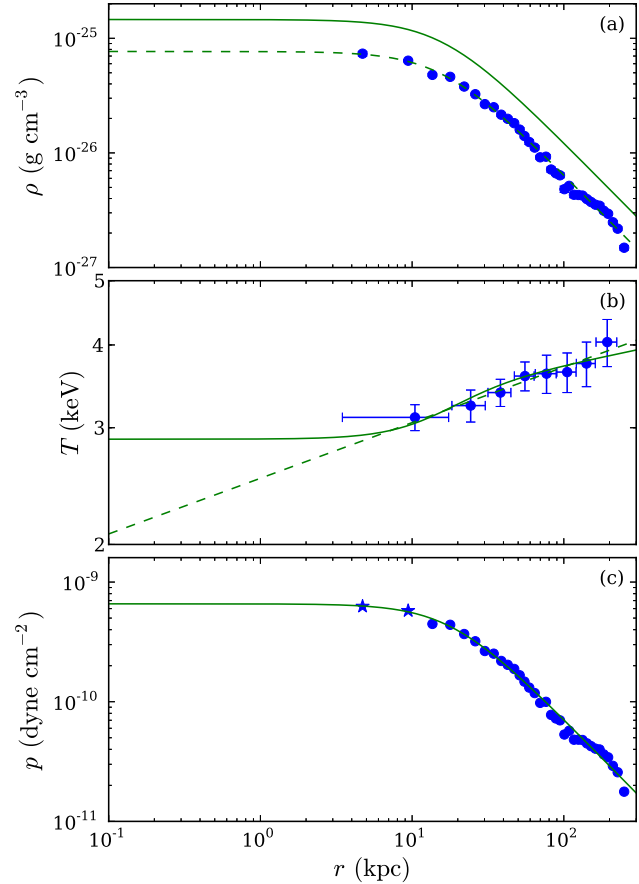}
\caption{Radial thermodynamic profiles for the Hydra A galaxy cluster. \textbf{(a)} Electron density (data points and dashed line), and corresponding total particle density (solid line) assuming a fully ionized plasma; Error bars are smaller than the points. \textbf{(b)} Electron temperature (data points and dashed line) and the temperature profile obtained from the total particle density and the pressure profiles (solid line); \textbf{(c)} Total pressure. The points in (a) and (b) are data for the electron density and electron temperature, respectively, obtained by \citet{david01} from X-ray observations of the Hydra A atmosphere. The dashed curves are fits to the data points using Eqn~\ref{density_profile} for (a) and a power-law for (b). The points in (c) were calculated from the density data points and the temperature fit. The stars represent the additional data points that we obtained through this method inside of 10 kpc. These two data points are important in constraining the profiles in the innermost region. The line in (c) is a fit to the data with Eqn.~\eqref{pressure_profile}.
Finally the temperature profile (solid line in panel b)) is obtained from the total particle density and the pressure profile. \newline
(A colour version of this figure is available in the online journal)}
\label{profiles}
\end{figure}

\begin{table}
\caption{Atmosphere profile parameters. Fits to data by \citet{david01} and extrapolation to $r<10$ kpc.} 
\centering
\begin{tabular}{r c c}
 \hline\hline
 & Parameter & Best fit value \\
 \hline
 \multirow{3}{*}{Density profile} & $r_{\rho 0}$ &  15.94 kpc \\
 & $\rho_0$     &  $1.49\times 10^{-25}$ g cm$^{-3}$ \\
 & $\alpha_{\rho}$ & 0.67  \\  
 \hline
 \multirow{2}{*}{Temperature profile} & $a_T$   & 5.66     \\
 & $b_T$   & $8.4\times10^{-2}$ \\
 \hline
 \multirow{3}{*}{Pressure profile} & $r_{p0}$  & 18.21 kpc	 \\
 & $p_0$     &	$6.58 \times 10^{-10}$ dyne cm$^{-2}$ \\
 & $\alpha_{p}$ & 0.65  \\
 \hline
\end{tabular}
\label{halo parameters}
\end{table}

In order to construct definitive simulations of the inner jet propagation, we require knowledge of the distribution of the ambient density and pressure on a 10~kpc scale. In this section we present useful analytical fits for the density, temperature, and pressure in the cluster environment, which we use to estimate the pressure and density in the inner 10 kpc.

We assume that Hydra A's atmosphere prior to the passage of the jet is hydrostatic, following a spherically-symmetric density distribution of the form
\begin{equation}
\rho(r) = \frac{\rho_0}{(1+r^2/r_{\rho0}^2)^{\alpha_{\rho}}}\:.
\label{density_profile}
\end{equation}
$\rho_0$, $r_{\rho0}$, and $\alpha_{\rho}$ are determined through a least-squares fit of the function given in Eqn.~\eqref{density_profile} to the models for the cluster density inferred from the X-ray surface brightness by\citet{david01}. The data and the fitted density profile are shown in Fig.~\ref{profiles} a). The pressure distribution of the atmosphere depends on the temperature distribution through $p = \rho k_B T/\mu m$, where $k_B$ is Boltzmann's constant, $\mu$ is the molecular weight and $m$ is the atomic mass unit. However, there is no observational data for the temperature inside of 10 kpc. We therefore use a power law temperature fit, $\log T = a_T+b_T\log r$, as shown in Fig.~\ref{profiles} b), to the \citet{david01} data and the density profile given by Eqn.~\eqref{density_profile} to obtain corresponding pressure values for two additional points within a radius of 10 kpc; these are distinguished from the other data points by the star-shaped symbols in Fig.~\ref{pressure_profile} c). The two additional extrapolated data points are important in constraining the shape of the flattening pressure profile toward the core of the galaxy. We adopt the following analytic expression for the pressure profile of the ICM
\begin{equation}
p(r) = \frac{p_0}{(1+r^2/r_{p0}^2)^{\alpha_p}}\:.
\label{pressure_profile}
\end{equation}
A least squares fit to the pressure data points is used to obtain the parameters  $p_0$, $r_{p0}$, and $\alpha_p$.  
 We then obtain the final temperature fit (solid line in panel b)) using the total particle density (solid line in panel a)) and the pressure profile (solid line in panel c)). 
 The best-fit parameters for the fits to the density, temperature, and pressure data are summarised in Table \ref{halo parameters}. 

For a hydrostatic environment we now have a gravitational acceleration profile
\begin{equation}
g(r) = - \frac {1}{\rho} \frac {dp}{dr} =  -2\alpha_p \frac{p_0}{\rho_0} \frac{r}{r_{p0}}\frac{(1+r^2/r_{\rho0}^2)^{\alpha_{\rho}}}{(1+r^2/r_{p0}^2)^{1+\alpha_p}}\:.
\end{equation}

%
%
\section{Code and Simulation Parameters} \label{s:code}


\begin{table*}
\caption{Simulation parameters. In all simulations, $P_{\rm{jet}}=10^{45} \rm \ erg \ s^{-1}$.}
\centering
\begin{tabular}{l * {9}{c}}
\hline \hline
Simulation  & $r_{\rm{jet}}(pc)$ & $p_{\rm{jet}}/p_{\rm{a}}$ & $\beta$ & $\chi$ & $\eta$ & $\phi$ (rad cm$^{-2}$) & $\Psi_{6\rm{cm}}$ (rad) & $\Psi_{20\rm{cm}}$ (rad) \\
\hline
	\hline
	 Ai    &  180 &   2  &  0.40  &     251.19 &   5.62$\times10^{-3}$  &    4.95$\times10^{-4}$         &     1.78$\times10^{-2}$		&  1.98$\times10^{-1}$      \\	 
  	 Aii 	& 180 &  	2  &  0.70 & 	23.17 &	 5.18$\times10^{-4}$ & 	4.56$\times10^{-5}$ 	&	1.64$\times10^{-3}$		&  1.83$\times10^{-2}$	  \\
	 Aiii 	& 180 &  	2  &  0.75 & 	15.03 &	 3.37$\times10^{-4}$ &	2.97$\times10^{-5}$ 	&	1.07$\times10^{-3}$		&  1.19$\times10^{-2}$	  \\
	Aiv 	& 180 & 	2  & 	0.80 & 	9.27   &  	2.07$\times10^{-4}$  &	1.83$\times10^{-5}$		&	6.57$\times10^{-4}$ 	&  7.30$\times10^{-3}$	 \\
	Av 	& 180 & 	2  & 	0.85 & 	 5.10  & 	1.14$\times10^{-4}$ &	 1.01$\times10^{-5}$	& 	3.62$\times10^{-4}$		&  4.02$\times10^{-3}$  	 \\
	 Avi 	& 180 & 	2  &	0.90 &  	2.14   & 	4.79$\times10^{-5}$  &	 4.22$\times10^{-6}$	&  	1.52$\times10^{-4}$		&  1.69$\times10^{-3}$ 	\\
	 Avii  &  180  &  2  & 0.95  &   0.11  &   2.40$\times10^{-6}$       &    2.11$\times10^{-7}$		&	7.60$\times10^{-6}$		&   8.45$\times10^{-5}$  \\
		\hline
        Bi    &  150  & 2  &  0.40  &   366.98  & 8.21$\times10^{-3}$  & 6.02$\times10^{-4}$ 	& 	2.17$\times10^{-2}$ 	& 	2.41$\times10^{-1}$  \\
  	 Bii 	& 150 &  2  &  0.70 &   34.90 & 7.80$\times10^{-4}$  &	5.73$\times10^{-5}$		&	2.06$\times10^{-3}$		&	2.29$\times10^{-2}$  \\
	 Biii 	& 150 &  2  &  0.75 &   23.00 & 5.14$\times10^{-4}$  &	3.78$\times10^{-5}$		&	1.36$\times10^{-3}$		&	1.51$\times10^{-2}$  \\
	Biv 	& 150 & 2  &  0.80 & 14.45   & 3.23$\times10^{-4}$   &	2.37$\times10^{-5}$		&	8.54$\times10^{-4}$ 	&  	9.49$\times10^{-3}$	 \\
	Bv 	& 150 & 2  &  0.85 &  8.28  & 1.85$\times10^{-4}$     &     1.36$\times10^{-5}$		&	4.89$\times10^{-4}$		&   	5.44$\times10^{-3}$	 \\
	Bvi 	& 150 & 2  &  0.90 & 3.87  &  8.64$\times10^{-4}$     &	6.34$\times10^{-6}$		&	2.28$\times10^{-4}$		& 	2.54$\times10^{-3}$ 	\\
	Bvii 	& 150 & 2  &  0.95  & 0.79 & 1.78$\times10^{-5}$	&    1.30$\times10^{-6}$		&      4.69$\times10^{-5}$		&      5.21$\times10^{-4}$   \\
 	Bviii    & 150 & 5 & 0.70 & 11.86  & 6.63$\times10^{-4}$      & 	4.87$\times10^{-5}$		& 	1.75$\times10^{-3}$		&      1.95$\times10^{-2}$   \\
	Bix    & 150 & 5 & 0.75 & 7.43   & 4.15$\times10^{-4}$      & 	3.05$\times10^{-5}$		&	1.10$\times10^{-3}$		&	1.22$\times10^{-2}$ \\
	Bx    & 150 & 5  & 0.80  &  4.28 & 2.39$\times10^{-4}$    &    1.76$\times10^{-5}$		&	 6.32$\times10^{-4}$	& 	7.02$\times10^{-3}$  	\\
	Bxi  & 150 & 5  &  0.85 &   2.04  &   1.14$\times10^{-4}$    &	8.39$\times10^{-6}$ 	&  	 3.02$\times10^{-4}$	&  	3.35$\times10^{-3}$ \\
        Bxii   &  150 & 5  &  0.90 &   0.48  &  2.70$\times10^{-5}$   &	1.98$\times10^{-6}$ 	&    	 7.13$\times10^{-5}$	&  	7.92$\times10^{-4}$	 \\
	\hline
        Ci     &  100 & 5  & 0.40    &  329.08  & 1.84$\times10^{-2}$ &   9.00$\times10^{-4}$		&      3.24$\times10^{-2}$		&      3.60$\times10^{-1}$ \\
	Cii	& 100  & 5  & 0.70  & 31.06 & 1.74$\times10^{-3}$	  &   8.50$\times10^{-5}$		&	3.06$\times10^{-3}$		&	3.40$\times10^{-2}$ \\
	Ciii	& 100  & 5  & 0.75  & 20.41 & 1.14$\times10^{-3}$ 	  &   5.58$\times10^{-5}$	 	&      2.01$\times10^{-3}$		&	2.23$\times10^{-2}$ \\
  	Civ    & 100 &  5  & 0.80  & 12.75  &  7.83$\times10^{-4}$  &   3.49$\times10^{-5}$  	&	1.26$\times10^{-3}$ 	&  	1.40$\times10^{-2}$ 	\\
	Cv   & 100 &  5  &  0.85 &   7.24   &  4.45$\times10^{-4}$  &   1.98$\times10^{-5}$ 		&  	7.13$\times10^{-4}$		&      7.92$\times10^{-3}$  \\
	Cvi   & 100 & 5  &  0.90 &   3.30  &  2.03$\times10^{-4}$   &	9.03$\times10^{-6}$  	&  	 3.25$\times10^{-4}$	&       3.61$\times10^{-3}$	 \\
	Cvii   & 100 & 5 & 0.95  & 0.57  &  3.18$\times10^{-5}$	  &    1.56$\times10^{-6}$		&      5.61$\times10^{-5}$		&       6.33$\times10^{-4}$  \\
	Cviii  & 100 &  5  &  0.96  & 0.15  & 8.33$\times10^{-6}$	  &    4.08$\times10^{-7}$		&      1.47$\times10^{-5}$		&      1.63$\times10^{-4}$ \\
	\hline
	 Di 	& 120  & 5   &  0.50    & 96.96	  & 5.48$\times10^{-3}$	 & 	3.22$\times10^{-4}$ 	&	1.16$\times10^{-2}$		&  1.29$\times10^{-1}$	  \\
	 Dii 	& 100  & 5   &  0.50 & 144.34  & 8.07$\times10^{-3}$ &	3.95$\times10^{-4}$         &	1.42$\times10^{-2}$		&  1.58$\times10^{-1}$	  \\ 
	 Diii 	& 80    & 10   &  0.50  & 111.14  &  1.24$\times10^{-2}$    &	4.86$\times10^{-4}$		&	1.75$\times10^{-2}$ 	&  1.95$\times10^{-1}$	 \\
	 Div 	& 80    & 15   &  0.50  &  71.60 	  & 1.20$\times10^{-2}$     &	 4.70$\times10^{-4}$	& 	1.69$\times10^{-2}$		&  1.88$\times10^{-1}$  	 \\
	 Dv 	& 60    & 10   &  0.50  &  203.38  & 2.27$\times10^{-2}$     &	 6.68$\times10^{-4}$	&  	2.40$\times10^{-2}$		&  2.67$\times10^{-1}$ 	\\
	 Dvi 	& 60    & 15   &  0.50  &  133.10  & 2.23$\times10^{-2}$     &	 6.55$\times10^{-4}$	&  	2.36$\times10^{-2}$		&  2.62$\times10^{-1}$ 	\\
	 \hline
\end{tabular}
\label{t:sim_par}
\end{table*}

For our simulations we use the the publicly available PLUTO code \citep{mignone07} and produce two dimensional axisymmetric hydrodynamic models of the jet-ICM interaction in Hydra A. Since our models involve relativistic velocities we use the relativistic hydrodynamic (RHD) module available in PLUTO. 

The $(r, z)$ computational domain for the two dimensional axisymmetric simulations is a cylinder of radius $r=25$ kpc and height $z=50$ kpc. Using a stretched grid we define a high resolution grid within the central $10\,\mathrm{kpc}\times1\,\mathrm{kpc}$ region, giving us 10 cells across the jet inlet, and a lower resolution in the outer regions. We impose an axisymmetric boundary condition for the boundary $r=0$, and a reflective boundary condition for $z=0$.  The remaining boundaries are set to outflowing boundaries.

We use the \citet{taub1948a} equation of state, a quadratic approximation to the exact Synge--J\"{u}ttner relativistic perfect gas equation of state \citep{juttner1911a,synge1957a}, which yields $\gamma\rightarrow5/3$ in the low temperature limit, and $\gamma\rightarrow4/3$ in the high temperature limit. Because the radiative cooling time of the ambient gas and the synchrotron cooling time of the jet plasma are large compared to the simulation time, we do not include radiative cooling in our simulations.

The initial conditions for the ambient medium representing the hot ICM surrounding Hydra A are the hydrostatic thermodynamic profiles found in \S~\ref{s:cluster}. 

To determine the optimal values for the three initial jet parameters $r_{\rm{jet}}$, $p_{\rm{jet}}$, and $\beta$ for the Hydra A northern jet, we compare the radius profile of the jet and the locations and spacing of the reconfinement shocks in our simulations with the observed radius profile and shock positions as indicated by the locations of the two bright knots. The thirty three sets of parameters that we have used are summarised in Table~\ref{t:sim_par}. We have not utilised every possible combination of parameters since we have restrictions on the jet radius minimum of 160 pc. We have not used models with five times over-pressured jet with jet inlet radius 180 pc and two times over-pressured jet with inlet radius 100 pc since they will produce much larger or smaller minimum in the radius profile than 160 pc. Since by experimenting models with lower jet velocities we obtain significantly large shock spacing compared to the observed shock spacing, we have not presented models with $\beta < 0.4$. A grid of models with jet $\beta = 0.5$ which exhibit larger shock spacings is presented in \S~\ref{s:b_r} 

Some derived parameters, namely the density parameter $\chi$, the density ratio $\eta$ of the jet and the atmosphere at the jet base, the rotation measure (RM) $\phi$, and the Faraday rotation angle $\Psi$ at 6 cm ($\Psi_\mathrm{6cm}$) and 20 cm ($\Psi_\mathrm{20cm}$) are also summarised in Table \ref{t:sim_par}. The rotation measure and Faraday rotation of the central jet with electron density $n_\mathrm{\mathrm e,jet}$(=$\rho_{\mathrm jet}$(1 + 2 $n_{\mathrm He}$/$n_{\mathrm H}$)/u(1 + 4$n_{\mathrm He}$/$n_{\mathrm H}$), where $u$ is an atomic mass unit), magnetic field along the line of sight $B_z$ (we use $35 \> \mu\,\mathrm{G}$, approximately the minimum energy magnetic field near the jet base), differential plasma depth $dl$, jet radius $R_{\rm{jet}}$, total plasma depth $L=2R_{\rm{jet}}$, and wavelength $\lambda$ are calculated from
\begin{eqnarray}
\phi &=& 8.1\int n_\mathrm{e,jet} B_z dl \quad \rm rad \, cm^{-2} \nonumber \\
&=& 8.1\times10^{-5}\, \left(n_{e, jet}\right) \left(\frac{B_z}{\mathrm{\mu G}}\right)  \left(\frac{2R_{\rm{jet}}}{\mathrm{kpc}} \right) \rm rad \, cm^{-2}
\end{eqnarray}
where the units of $B_z$ and $l$ are Gauss and cm, respectively. The total Faraday rotation through the jet is given by:
\begin{equation}
\Psi_{\rm rad} = \phi \lambda^2 \:.
\end{equation}

We calculate these quantities as an additional check to ensure that our jet parameters are consistent with the observation that the radio emission along the length of the jet is polarized. The internal Faraday rotation should be much less than unity for consistency between our models and the observations. Note however, that the values given in Table~\ref{t:sim_par} are maximum values and do not take into account the angle between the magnetic field and the line of sight, the possibility that the magnetic field strength may be below equipartition, or the occurrence of field reversals. Nevertheless, all of the Faraday rotation values are comfortably less than unity and in our best models, Ciii, Civ and Cv, much less than unity.

We group our runs into four sets as set out in Table~\ref{t:sim_par}; Sets A, B and C correspond to simulations with initial jet radii of $0.18  \> \rm kpc$, $0.15 \> \rm kpc$ and $0.10 \> \rm kpc$, respectively. Set D corresponds to model with jet $\beta = 0.5$ and initial jet radii $0.12, \> 0.10 \> 0.08 \> \rm and \> 0.06 \> kpc$   

%
%

\section{Simulation Results}\label{s:sims}

In this section we present the results of our two-dimensional axisymmetric hydrodynamic simulations, including the parameter study described above. We have conducted a series of simulations to cover the parameter space described in 
Table~\ref{t:sim_par}. We first describe the results of our parameter space study, which enable us to constrain the jet velocity and other jet parameters at 0.5 kpc from the black hole. These provide best fit models for the northern Hydra A jet. Using one of the best fit models, Civ, we then discuss the association of biconical shocks with the bright knots, the turbulent transition of the jet, and the flux density ratio between the northern and southern jet of Hydra A. Finally, based on the discrepancy between the simulated and the observed flux density ratio, we explore the possibility of varying the angle of inclination within the range defined by \citet{taylor93}. 

%
%

\subsection{Parameter space study for the northern jet}\label{s:param_study}

The aim of our parameter space study is to obtain optimal values for the jet parameters, in particular, the jet radius, the jet pressure and the jet velocity at $0.5$ kpc from the core.

\begin{figure}
\includegraphics[width=\linewidth]{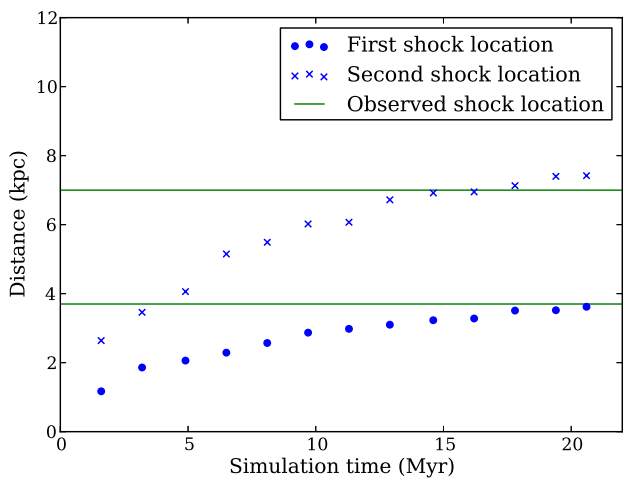}
\caption{
Evolution of the locations of the first (blue dots) and second (blue crosses) shocks with time for run Civ. The horizontal lines represent the observed shock locations. This figure shows that the first two reconfinement shocks move downstream with time and asymptote towards  3.6 and 7.4 kpc at approximately 20 Myr. 
\newline
(A colour version of this figure is available in the online journal)}
\label{f:s_ev}
\end{figure}

\begin{figure*}
\includegraphics[width=\textwidth]{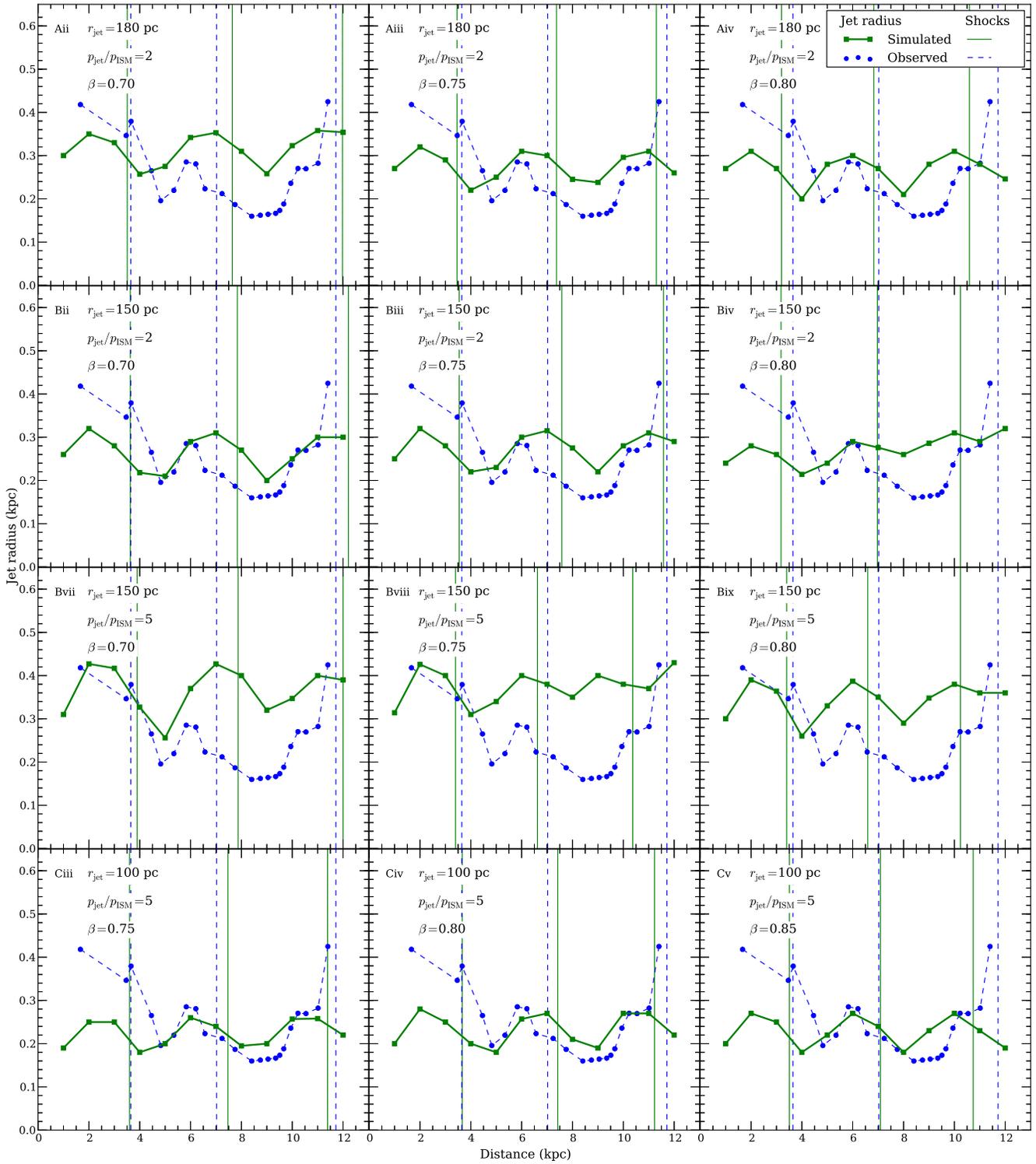}
\caption{
Jet radius profiles and shock positions along the jet extracted from selected hydrodynamic simulations. The green line with squares, and the blue lines with circles represent simulated and observation data of radius, respectively. The blue dashed and green solid vertical lines represent the observed and simulated shock locations, respectively. The top row of panels are simulations from set A, the second and third rows of panels are simulations from set B, and the bottom row of panels are simulations from set C. In the simulations shown in the upper two rows of panels, $p_{\rm jet}/p_{\rm ICM}=2$, whereas in those shown in the lower two rows of panels, $p_{\rm jet}/p_{\rm ICM}=2$. The left, middle, and right column of panels, show simulations for which $\beta=0.75$, 0.80, and 0.85, respectively. A visual comparison of the jet radius profiles and shock positions between the simulations and observations shows that, of our models, Ciii, Civ and Cv give the good fit models. 
\newline
(A colour version of this figure is available in the online journal)}
\label{f:parameter_study}
\end{figure*}

As discussed in \S~\ref{s:model}, the natural wavelength for the occurrence of reconfinement shocks in a supersonic jet is directly related to the jet velocity. We vary the jet velocity, at the same time consistently varying the density parameter $\chi$ to maintain a constant jet kinetic power, noting the location of the first two reconfinement shocks in the jet for each run.
As the cocoon pressure decreases with increasing size the locations of the reconfinement shocks of each run evolve with time. The shocks gradually shift downstream and reach asymptotic values at approximately 20 Myr. We take these asymptotes as the location of the shocks.  Figure~\ref{f:s_ev} shows the evolution of the location of the first (blue dots) and second (blue crosses) shocks with time and the observed location of the shocks (green lines) for run Civ.

The shock positions also vary on a short time scale, oscillating about a mean position.

These variations occur because the pressure field in the backflow adjacent to the jet changes intermittently as a result of the turbulence in the cocoon. Hence, for each run, we have measured the position of the jet shock at five time steps separated by 100 kyr in time. Figure~\ref{f:parameter_study} shows the jet radius profiles and reconfinement shock positions from selected simulations. The results from simulations in set A, B, and C are shown in the top, middle two, and bottom rows, respectively. We compare the simulated shock positions (solid vertical lines) with the observed shocks in Hydra A (dashed vertical lines) and  also compare the simulated jet radius profiles (solid green lines and squares) with the observed jet radius profile (solid blue line and circles).

In assessing these models, one first notes a strong dependence of shock location on jet speed, as expected, and we use this as the first discriminant in selecting candidate best fit models. This narrows the choice to Aiii, Biii, Ciii, Civ, and Cv. Then, focusing on the radius profile, in models Aii, Bii, the jet radius does not contract sufficiently at large distances, which make these two models less appealing. At the same time, we note that the remaining models Cii, Ciii and Civ provide poor radius fits within 3~kpc.  However, the first three data points are derived from a region, which is affected by the emission from the core \citep[see][Fig. 3]{taylor90}. It is also possible that our models do not capture the details of the initial jet-ISM interaction in this region.

Hence, we concentrate on the data points further out from the core. Consequently our choice for the best fit models are Ciii, Civ and Cv. Our preference for these three models is based on the fact that the simulated radius shows larger excursions between minima and maxima as exhibited by the data. The parameters for our best fit models Ciii, Civ and Cv are $r_{\rm jet} = 100 \rm pc$, $p_{\rm jet}/P_{\rm ISM} = 5$ and $\beta = 0.75, 0.80, \rm \ and \ 0.85$, respectively. We also note that the last point in the observed radius profile jumps significantly. We attribute this to the onset of turbulence in the jet where it makes a transition to a plume. The third knot/shock may be affected by this transition so that in deciding between models we have mainly concentrated on the first two knots. 

%
%
\subsection{The surface brightness of the knots in the Northern Jet}
\label{s:knot}

\begin{figure}
\centering
\includegraphics[width=\linewidth]{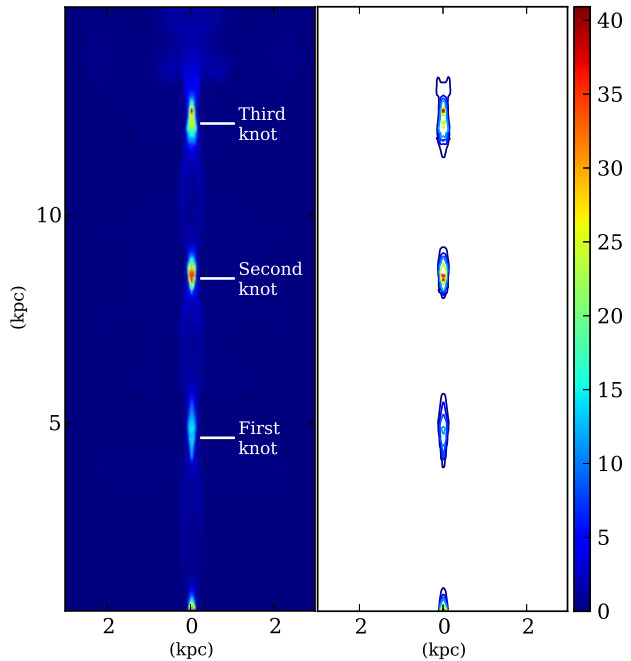}
\caption{Synthetic surface brightness for model Civ based on an emissivity $j_{\nu} = \delta^{(2+\alpha)} p^{(3+\alpha)/2} \nu^{-\alpha}$, where $\delta= 1/\Gamma(1-\beta \cos 42^\circ)$ is the Doppler factor. The right panel shows the surface brightness contours of the left panel. The contour levels are 4, 8, 11, 14, 25, 35, 40 in arbitrary units. \newline (A colour version of this figure is available in the online journal)}
\label{radio_morphology}
\end{figure}

To strengthen the association of biconical shocks with the bright knots in the Hydra A northern jet we present a synthetic radio image of one of our best fit models, Ciii, based on an assumed synchrotron emissivity $j_{\nu} \approx {\psi} \, \delta^{2+\alpha} \, p^{(3+\alpha)/2}$, where $\psi$ is the relativistic gas tracer, the Doppler factor $\delta= 1/\Gamma(1-\beta \cos 42^\circ)$ and the pressure dependence assumes that the magnetic pressure is proportional the non-thermal particle pressure \citep[see][\S 5.4]{sutherland07a}. Integrated along rays
 $I_\nu = \int j_\nu ds$,
 this emissivity provides a semi-quantitative estimate of the surface brightness corresponding to this model.

Fig.~\ref{radio_morphology} (left panel) shows the synthetic surface brightness of the simulated jet. The contour image of the synthetic surface brightness is shown in the right panel. Here we see that, in the shocked zone beyond each biconical shock,  the pressure increases, producing bright knots in each region. This image reproduces some qualitative features of the data: The second and third knots are significantly brighter and more extended than the first knot. However, the 
brightness ratios of the knots are not reproduced. 
Observationally (corrected for resolution) the second knot is 8.7 times brighter than the first and the third knot is 3 times brighter than the second. The model values are 2.5 and 1.14 respectively. In addition, in the observed jet, the FWHM extent of the second knot in the jet direction is 3.3~kpc compared to 0.6~kpc for the model. These differences may possibly be attributed to the approximate magnetic field model, which we have used, or the lack of turbulent three dimensional structure in our simulations. These are aspects to which we can return with three-dimensional simulations with magnetic field. 

%
%
\subsection{Transition to turbulence}
\begin{figure*}
\centering
\includegraphics[width=\textwidth]{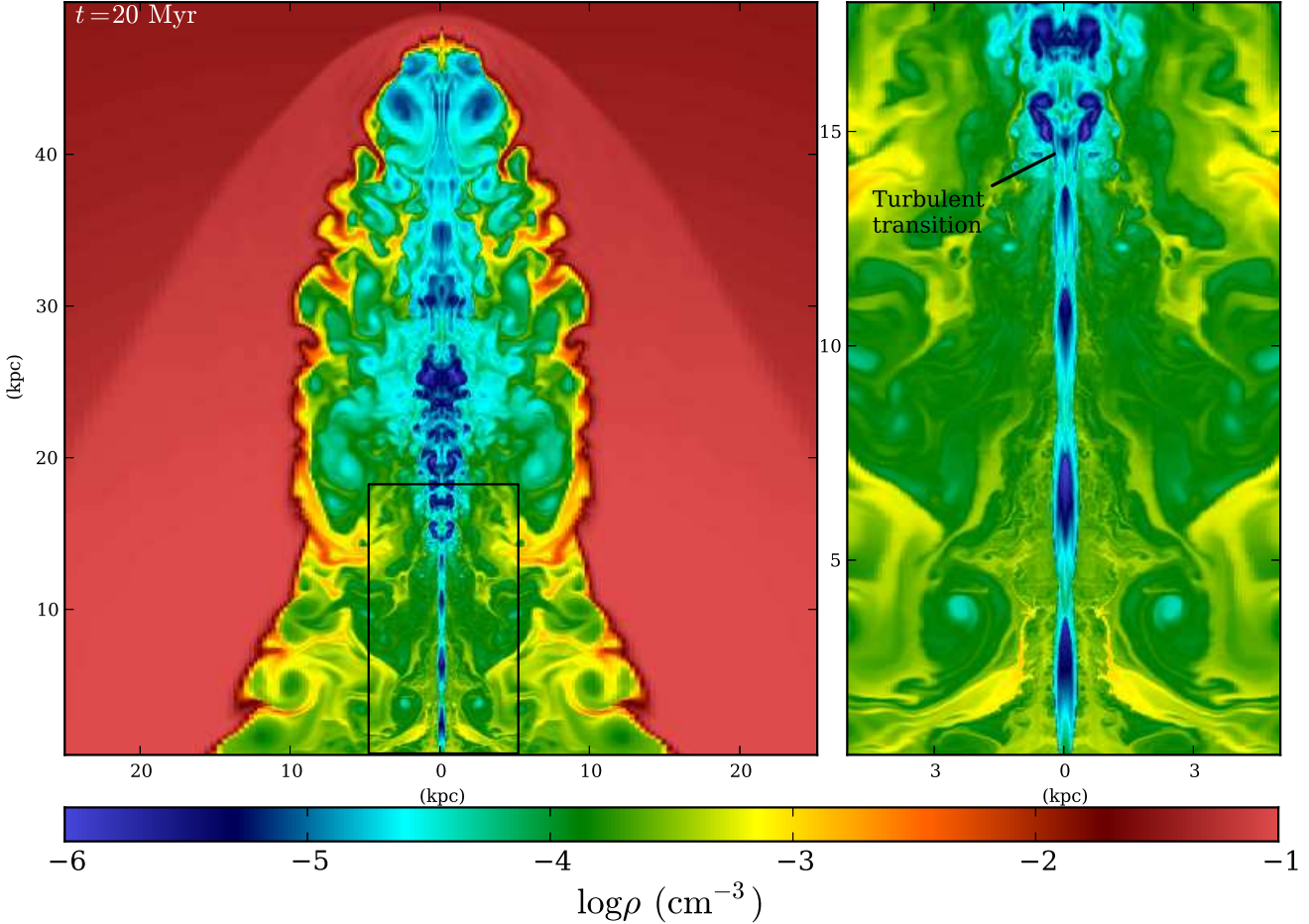}
\caption{Logarithmic density snapshot for run Civ at $t=20\,\mathrm{Myr}$ in the left panel. The right panel shows the zoomed in central zone marked with black rectangle in the left panel. This is one of our best fit models, which yields the correct location of the first two biconical reconfinement shocks in the northern jet of Hydra A. A transition to turbulence occurs due to significant shock deceleration of the jet in the reconfinement shocks and the developing Kelvin-Helmholtz instability. 
\newline (A colour version of this figure is available in the online journal)}
\label{t_trans}
\end{figure*}
 
 \begin{figure}
\centering
\includegraphics[width=\linewidth]{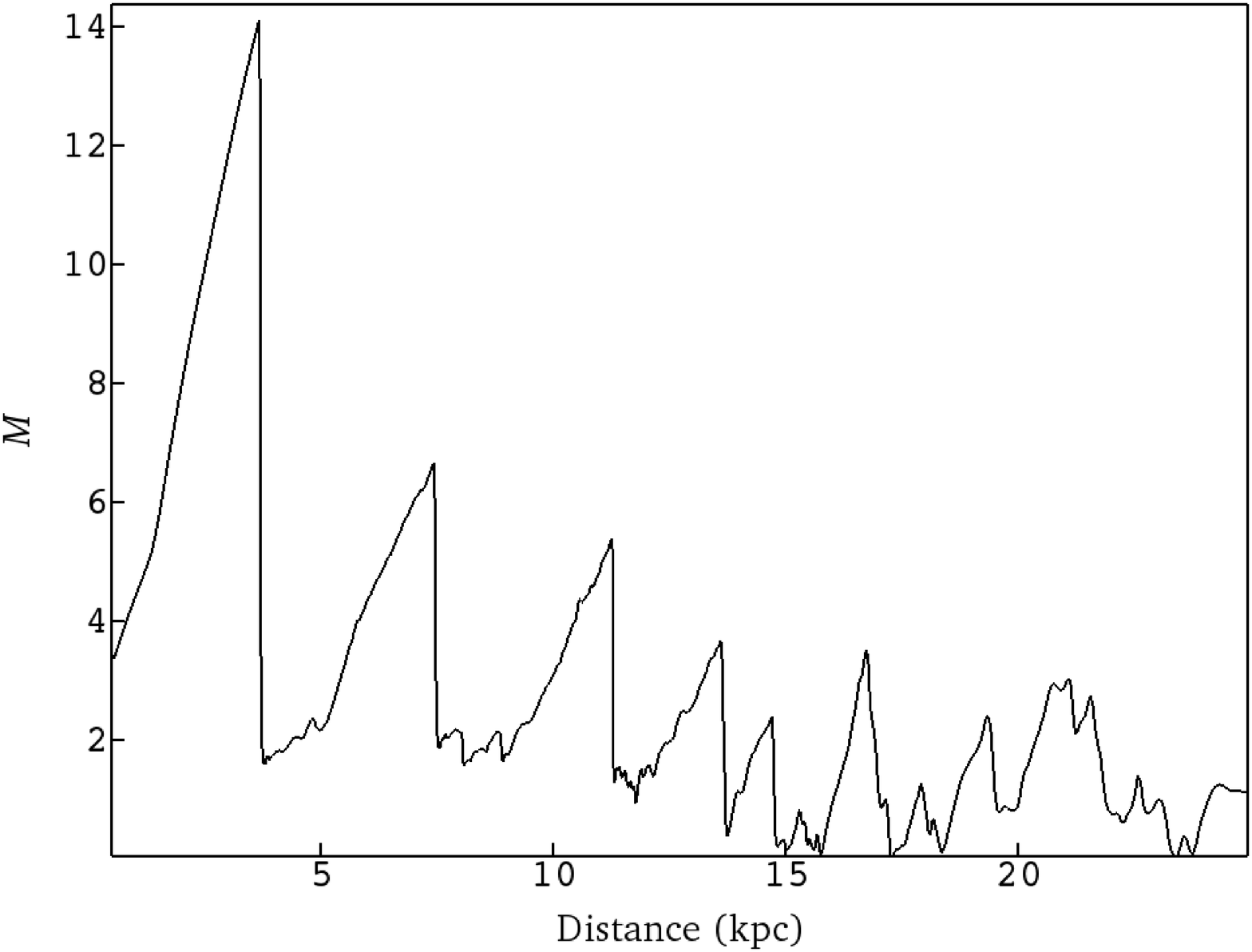}
\caption{The Mach number of the jet of model Civ at different locations along the jet axis.}
\label{mach}
\end{figure}

These two-dimensional models cannot adequately reproduce the structure of the entire source, in particular the plume like regions beyond approximately $7.4^{\prime\prime}$. These are probably the result of three-dimensional turbulence and/or precession, and these effects will be addressed in subsequent papers in this series. However, we note that our numerical models \emph{qualitatively} reproduce the turbulent transition of the jets to plumes, albeit at a distance of 14~kpc compared to approximately 11 kpc deprojected in Hydra~A. In the density image snapshot at approximately 20 Myr of run Civ, Fig.~\ref{t_trans} (the left panel shows the full computational domain and the right panel is the zoom in section indicates by the rectangle in the left panel), a series of biconical  shocks appears in the jet.  Deceleration of the jet occurs at these shocks and the jet becomes subsonic after the fourth shock at $\sim 14 \rm \ kpc$ (see the variation of Mach number of the flow with distance along the jet axis in Fig.~\ref{mach}). Beyond 14~kpc the jet transitions to turbulence as a result of the axisymmetric Kelvin-Helmholtz instability, which becomes stronger as the Mach number decreases.) 
Although our axisymmetric jet simulations shed some light on the turbulent transition of the jet, it is well known that turbulence and the formation of plumes are three dimensional phenomena, especially in supersonic flows. We study the details of these features of the inner 50 kpc of the Hydra A jets in our ensuing three dimensional study.

%
%

\subsection{Brightness ratio of the jets} \label{s:b_r}
 \begin{figure*}
\centering
\includegraphics[width=\textwidth]{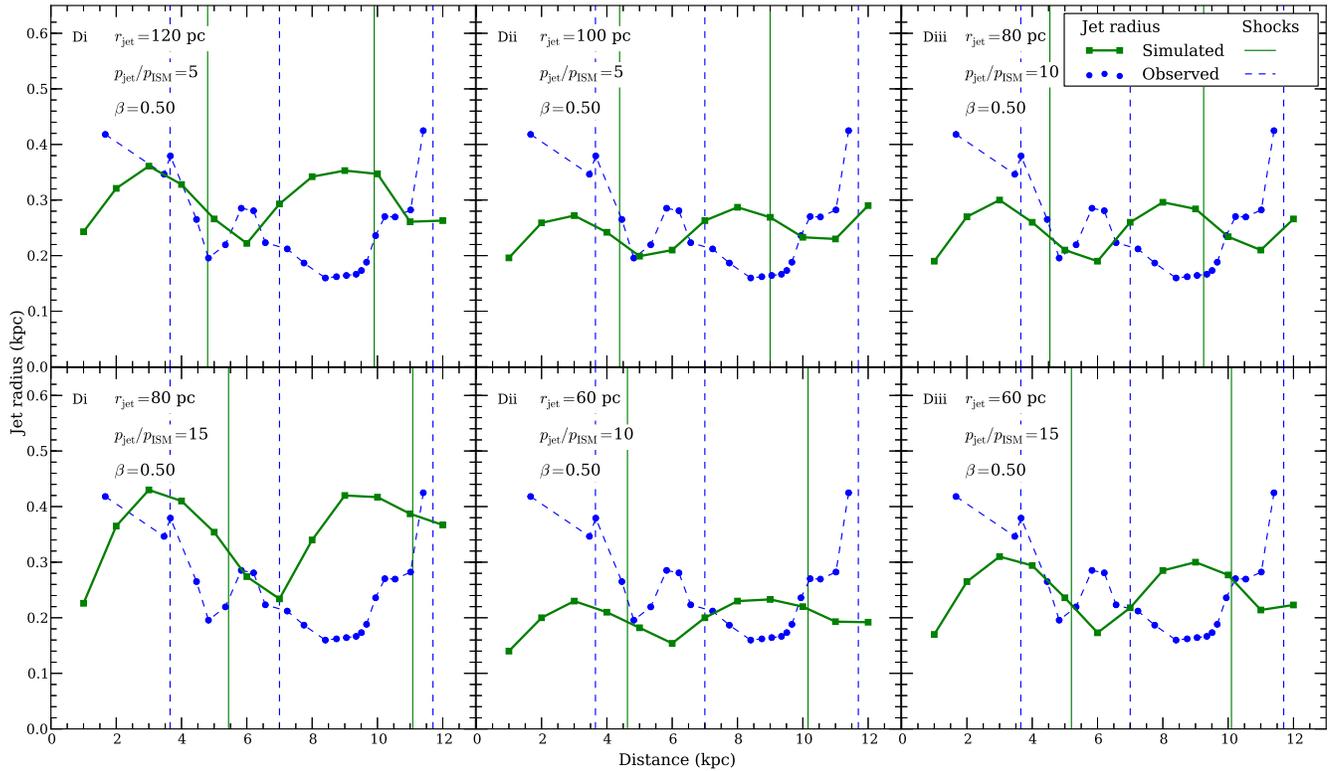}
\caption{Jet radius profile and shock positions along the jet for jet velocity 0.5. The green line with squares, and the blue lines with circles represent simulated and observation data of radius, respectively. The blue dashed and green solid vertical lines represent the observed and simulated shock locations, respectively.  \newline (A colour version of this figure is available in the online journal)}
\label{f:p_s_b5}
\end{figure*}

We have used the 6cm VLA data of \citet{taylor90} to determine the flux density ratio of the northern and southern jets within the first 10 kpc, obtaining  a value $\approx 7.0$\footnote{\citet{taylor90} quote a value of 1.9, which is close to the observed ratio within 1~kpc.} Attributing this ratio to Doppler beaming, and using the inclination estimated by \citet{taylor90}, implies a moderately relativistic jet  $\beta \approx  0.5$. However, our parameter space study produces a higher jet velocity $\sim 0.8$ which, on the basis of a simple estimate, would give a brightness ratio $\approx 40$. However, in our model, the emissivity is dominated by the decelerated post-shock regions of the jet, so that we estimate the brightness ratio from the synthetic brightness images of approaching and receding jets. With this approach, we obtain a simulated flux density ratio of 33 which still differs significantly from the observed value by a factor $\approx 5$.

We ran several additional models with jet $\beta = 0.5$,  different jet inlet radii 120~pc, 100~pc, 80~pc and 60~pc and different pressure ratio 5, 10, and 15, keeping the jet kinetic power constant at $10^{45} \rm \ ergs \ s^{-1}$. We have not decreased the jet radius below 60~pc because that would require an even more highly over pressured jet to obtain the correct radius profile. These $\beta=0.5$ models are summarised in Table~\ref{t:sim_par} (set D) and the comparison of the simulated and observed shock positions and radius profiles are shown in Fig.~\ref{f:p_s_b5}. It is evident that no model with the given jet kinetic power and jet $\beta = 0.5$ is able to produce good fits for both the shock position and the jet radius. The shock spacings are all significantly larger than the observed shock spacing and the radius profiles are mismatched with these models. 

From the above we can say that if we fix the inclination angle at the \citet{taylor90} value of $42^\circ$ and fix the jet kinetic power at $10^{45} \rm \ ergs \ s^{-1}$, then the jet pressure, jet velocity and the inlet jet radius at 0.5~kpc away from the core of the Hydra A northern jet are well-constrained by both the jet radius profile and the first two knot/shock spacings. The best-fit values are $\beta = 0.75 - 0.85$ and $r_j = 100 \> \rm pc$. Thus, there is a discrepancy in the flux density ratio between the simulated and observed jets. Two potential explanations of the low flux density ratio are: i) Since our models do not include the magnetic field, we employ the assumption $p\propto B^2/8\pi$ which gives a brightness ratio 33. If we further assume that the magnetic field is 2.5 times stronger in the southern jet of Hydra A we would obtain a lower brightness ratio $\sim 7$. ii) The southern jet is more dissipative since it is more twisted and produces more shocks producing a larger intrinsic emissivity than the northern jet.

%
%
\subsection{Variation of the inclination angle}\label{s:theta}

 \begin{figure}
\centering
\includegraphics[width=\linewidth]{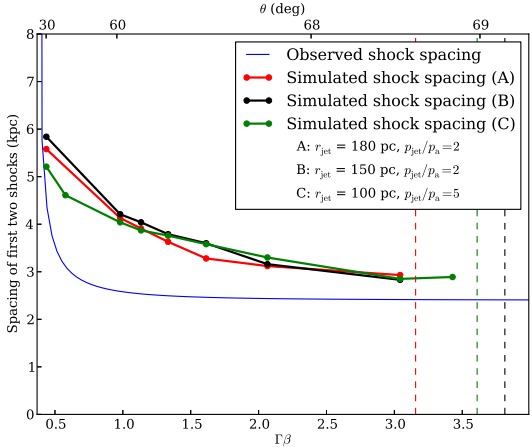}
\caption{Comparison of the observed  spacing  of the first two shocks (blue curves) with the corresponding simulated shock spacing (red points for model set A, black points for model set B, and green points for model set C) as a function of the 4-velocity $\Gamma \beta$. The dashed vertical lines represent the upper limits of $\Gamma \beta$ for each model (set A -- red, set B -- black, and set C -- green). These limits are estimated for $\chi = 0$ using Eqn.~(\ref{chi2}). \newline (A colour version of this figure is available in the online journal)}
\label{f:ss}
\end{figure}

In the above models we have used the angle between the jet and the line of sight, $\theta \approx 42^\circ$, estimated by \citet{taylor93} from the rotation measure asymmetry of Hydra~A. However, there is a fairly large uncertainty in their estimate of $\theta$ with $30^\circ \la \theta \la 60^\circ$. Increasing $\theta$ from $42^{\circ}$, would reduce the brightness ratio? However, for larger inclinations, the deprojected knot separation would decrease, and as we have seen with the above models, this would require a higher velocity than $0.8 \rm c$, tending to \emph{increase} the brightness ratio. Similar considerations apply if we \emph{decrease} the inclination. Nevertheless, is it possible that notwithstanding opposing effects, an inclination angle within the \citet{taylor90} range, and a jet velocity, can be found that are consistent with the dual constraints of surface brightness ratio and projected knot spacing? 

In order to assess this possibility we adopted the following procedure: For $\beta$ within the range, $0.35 < \beta < 0.98$ (the lower limit being defined by the brightness ratio, $R=7$) we first estimate the value of $\theta$ corresponding to $R = 7$, using the standard Doppler beaming formula, $\theta (\beta)= \beta^{-1}(R^{1/2.7}-1)(R^{1/2.7}+1)^{-1}$. For these values of $\theta(\beta)$ we determine the deprojected spacing between the first two shocks $D_{1,2}(\beta) = 2.25 /\sin \theta(\beta) \> \rm kpc$ given the observed spacing of 2.25~kpc. This is shown, as a function of the 4-velocity, $\Gamma \beta$, as the blue curve in Fig.~\ref{f:ss}. We then compare the observed deprojected knot spacings with the values inferred from the simulations so that in Fig.~\ref{f:ss} the simulated shock spacing, for model sets A, B and C, are also plotted as functions of $\Gamma \beta$. The upper limits on the 4-velocity for each model (estimated from Eqn.~(\ref{chi2})), associated with a zero density parameter, $\chi = 0$, are also shown as dashed vertical lines.

The first point to note with this comparison is that for most of allowable range of $\beta$ Fig.~\ref{f:ss} shows that the calculated shock spacing exceeds the observed, deprojected value. At the upper end of the $\beta$ range the simulated shock spacings for each model asymptote to $\approx 2.85$ kpc for values of $\Gamma \beta \ga 3$, i.e., $\beta \ga 0.95$. However the asymptote of the observed shock spacing $\approx 2.4$ kpc. Hence, there is an offset of approximately 0.5~kpc between the asymptotes of the simulated and observed shock spacing for $\Gamma \beta \ga 3$.  

At the other end of the allowable range of velocity, $\beta \approx 0.345$ ($\Gamma \beta = 0.368$), it could be inferred that the simulations and observations intersect at approximately this limiting value. However, this is the result of the steepness at $\beta \approx 0.345$ of the (blue) curve representing the observed deprojected shock spacing as a function of 4-velocity, rather than a real physical correspondence between observed and simulated values. It would be fortuitous if the jet initial velocity were to be almost exactly the same as the lower limit on the jet velocity implied by beaming. Hence we reject a solution at this end of the $\beta$ range on the basis of the ``fine-tuning'' that would be involved in accepting it. Another unappealing feature of a low-$\beta$ solution is that the jet would be initially heavy with $\chi \ga 300$. 
As we noted above, observations and modelling of X-ray observations of the lobes of radio galaxies indicate that jets are initially electron-positron in composition \citep{croston05a,croston14a} and $\chi \ga 300$ is inconsistent with this. 
 
Another way of looking at the issue of reconciling shock spacing and flux ratios is the following: Consider the simulation points near the upper end of the $\beta$ range in Fig.~\ref{f:ss}, where the discrepancy between the observed jet and simulated jets with $\chi \sim 1$ is the least. By way of example, consider the (green) point in simulation series C with $\beta=0.95$ ($\Gamma \beta = 3.04$). The simulated flux ratio (see \S~\ref{s:b_r}) for this model is 26.5, a factor of 3.8 higher than the observed value. Thus, even for these models there is an implication of intrinsic differences in the northern and southern jet rest-frame emissivities. Moreover, this ratio is not very different from the value of 33 for the $\beta =0.8$, $\theta = 42^\circ$ model considered earlier.

In view of the above, we conclude that, taking into account the modelling of shock spacing, radius evolution and surface brightness ratios, the most likely situation is that of fast, $\beta \ga 0.8$, jets with an intrinsic difference between the rest-frame emissivities of northern and southern jets.

%
%

\section{Summary and Discussion} \label{s:discussion}

The main thrust of this paper has been to understand the physics of the inner jets in Hydra~A with a view to using inferred parameters such as jet energy flux, pressure, density and velocity in large scale models of the radio source and its interaction with the cluster atmosphere.

We have focused on the following key features of the inner ($< 10 \> \rm kpc$ from the core): i) the bright knots in the northern jet at $\sim 3.7 \rm \ , \ 7.0 \ and \ 11.7$ kpc from the black hole ii) and the wave-like boundary of the northern jet. To this end, we have performed a series of two dimensional axisymmetric relativistic hydrodynamic simulations of the interaction of the northern Hydra A jet with the interstellar medium, particularly within the central $10 \> \rm kpc$.

To ensure that we use reasonable values for the jet parameters in our simulations, we have estimated the powers associated with the inner X-ray cavities of Hydra~A corresponding to the inner radio lobes. We have used 4.6~GHz radio observations by \citet{taylor90} to estimate the inner cavity power and have compared them with the estimates of \citet{wise07} for the same cavities based on the X-ray data. We obtain powers for the northern and southern cavities $\approx 1.8\times 10^{44} \> \rm ergs \> s^{-1}$ and $2.0\times 10^{44} \> \rm ergs \> s^{-1}$ respectively. These estimates are consistent with the \citet{wise07} estimates $\sim 2 \times 10^{44} \ \rm ergs \ s^{-1}$ for both cavities.
Hence, we adopt the total jet power obtained by \citet{wise07} $P_{\rm jet}=10^{45}$ erg $s^{-1}$ as our value for the jet power in the numerical models. Other jet parameters, the jet pressure $p_{\rm jet}$ ($= 2 p_{\rm a} \rm \ and \ 5  \ p_{\rm a}$; $p_{\rm a} = $ ambient pressure) and the jet inlet radius $r_{\rm jet}$($=180, 150, \rm \ and \ 100~pc$) are chosen based on the 23~cm VLBA and 6~cm VLA data of Hydra A \citep{taylor90}.

On the basis of the minimum pressure estimates we conclude that, in the lobes, $k$, the ratio of energy in other particles to that in relativistic electrons $\sim 10$. Moderate values of this parameter are supported by other recent studies: \citet{birzan08} estimated $k$ for a group of radio galaxies assuming that the radio lobes are in pressure equilibrium with the ambient medium. Their estimates include the Hydra A radio lobes at 1.4 GHz for which they obtained a value of $k \approx 13$. \citet{hardcastle10} studied the inverse-Compton X-ray emission from the outer Hydra A radio lobes and obtained values of $k \sim 17$ and 23 for minimum Lorentz factor cut-offs of $\gamma_1=1$ and 10 respectively. These estimates are all comparable given the different techniques used to derive them. 

For the X-ray atmosphere used in our simulations, we have constructed hydrostatic profiles surrounding the Hydra A radio source by fitting and extrapolating the density and temperature data from the X-ray observations of \citet{david01}.

The results of our numerical models of the interaction of initially conical and ballistic jet with the ambient medium support the idea that the consecutive biconical shocks are responsible for the bright knots in the northern jets of Hydra A.
With appropriate values of the initial jet pressure ratio and velocity the observed knot spacings and variation in jet radius are reproduced along a considerable section of the jet. We did not model the Southern jet since it is more twisted and a straight jet model would be inappropriate; furthermore the radius of the southern jet as a function of distance from the core has not been observationally determined.

From our comprehensive parameter study we have selected models Ciii, Civ and Cv as are our best fit models for the inner $\sim 10$ kpc radio structure of the northern jet. These jet models with initially conical and ballistic jet and over-pressured with respect to the environment by a factor of 5 produce four successive biconical reconfinement shocks before the jets become fully turbulent. The location of the first three shocks and the radius profile of the jet along its propagation closely match the location of the southern edge of the bright knots and the radius profile of the Hydra A northern jet. Constructing a synthetic surface brightness image we have shown that the biconical shocks produced in the simulated jet are associated with bright knots. For our best fit models of the northern jet, the jet parameters are $P_{\rm jet}=10^{45} \> \rm erg \rm s^{-1}$,  $r_{\rm jet}=100 \rm \> pc$, $p_{\rm jet}p_{\rm a} = 5$,  $\chi = 20.41, 12.75, 7.24$ and $\beta = 0.75, 0.8 \rm \ and 0.85$. Our estimated jet velocity for the northern jet of Hydra A $\approx 0.8 \rm \> c$ is consistent with a recent theoretical estimate of jet velocity for FRI radio jets provided by \citep{laing14}. In the course of modelling the surface brightness of 10 FRI radio sources they estimated a kpc scale jet velocity $\approx 0.8 \rm \> c$.

The brightness of the knots in our best fit model gradually increase with distance from the core, in a way that is qualitatively consistent with the observed jet. However the brightness ratio between the second and first knot and between the third and second knot for the simulated jet (run Ciii) $\approx 2.5$ and 1.14 respectively, differs from the observed brightness ratio $\sim8.7$ and $\sim3$. This discrepancy may arise as a result of the magnetic field increasing faster than the pressure along the jet and hence the assumption of $B^2/8\pi \propto p_{\rm jet}$ in the emissivity model would underestimate the emissivity increase along the jet. 

Our inferred relativistic jet velocity $\approx 0.8$ differs from the estimate based on the Doppler beaming $\approx 0.5$. Consequently we estimate a large flux density ratio 33 of the approaching and receding jet compared to the observed flux density ratio of 7. Our additional parameter study in \S~\ref{s:b_r} shows that the combination $\beta = 0.5$, jet kinetic power $10^{45} \rm \ erg \ s^{-1}$ and an inclination angle $\theta=42^\circ$ is unable to produce the correct shock locations and the profile of the jet boundary for any feasible combination of the jet inlet radius and pressure. Hence, one possibility is to adopt $\beta= 0.8$ and to attribute the different flux density ratio to a difference in intrinsic rest-frame emissivities. For example, the flux density ratio may be overestimated in our best fit model since we assume that the magnetic field is the same in both jets. If we assume that the magnetic field is 2.5 times stronger in the southern jet, the flux density ratio would be 7. Another possibility is that the observed value of the flux density ratio is low since the southern jet is more dissipative as a result of its greater bending and the greater number of shocks. 

Another possibility for the discrepancy between estimated and measured flux density ratios is that the angle, $\theta$, between the jet and the line of sight, inferred from the rotation measure asymmetry \citep[see][]{taylor93} differs from $42^\circ$. This is certainly possible given the range $30^\circ \la \theta \la 60^\circ$  estimated by \citet{taylor93}. Hence, we have used the jet velocity as a parameter, calculated the inclination required to give a northern to southern flux ratio of 7, calculated the deprojected spacing between the first and second knots and compared this with the simulated spacing. The result of this comparison has been that the simulated and observed spacings do not agree except at the lowest possible jet velocities, consistent with a beaming interpretation, $\beta \approx 0.35$. We have argued that a solution for the jet velocity at around $\beta = 0.35$ is unappealing since it is unlikely that the optimal velocity for knot spacing would be close to the lower limit from beaming. 

We conclude that the jet velocities $\ga 0.8 \, c$ and that there is an intrinsic asymmetry between the rest-frame emissivities of the northern and southern jets. This may be a result of different magnetic fields (by about a factor of 2.5) or higher dissipation in the southern jet.

The initial value (at 0.5~kpc) of the density parameter $\chi = \rho c^2 / 4 p$ derived from our simulations is also of interest for the parsec-scale value of this parameter. Assuming that the key has constant velocity from the pc-scale outwards, $\rho \propto r_{\rm jet}^{-2}$ and $p \propto r_{\rm jet}^{-8/3}$ so that $\chi \propto r_{\rm jet}^{2/3}$. From the VLBI images of \citet{taylor96} $r_{\rm jet} \approx 1 \> \rm pc$ in the 15.4~GHz image. Hence our best fit value of $\chi = 12.75$ extrapolates to 0.59 -- consistent with an electron-positron jet with $\gamma_1 \sim 1$ or an electron proton jet with $\gamma_1 \sim 700$.

Our conclusions are subject to the assumption of a low magnetic pressure in the jet and we have provided some justification for this assumption, on the sub-parsec scale in \S~3 and the lack of magnetic collimation from the parsec to kiloparsec scale. Nevertheless, the magnetic field evolves along a jet, and its downstream strength and influence on the dynamics is an interesting issue. Moreover, the magnetic field is important for the calculation of synchrotron emission so that even if it passively transported, its evolution is important for the calculation of surface brightness. Hence, the inclusion of a magnetic field in future simulations is of interest. However, as \citet{spruit11a} has shown there is a lot more physics to consider in this case, in particular the modelling of reconnection of three-dimensional magnetic field. Thus, while magnetic effects are important to consider in future work, their consideration is well beyond the scope of this paper, which we consider to be a useful first step in modelling features such as shock spacing and radial oscillations in order to estimate jet velocities.

The next paper in this series will focus on three dimensional modelling of the Hydra A jets and the details of the transition of the collimated jets to turbulent plumes.

\section*{Acknowledgments}

This research was supported by the Australian Research Council Discovery Project, DP140103341 \emph{The key role of black holes in galaxy formation}.  The computations were undertaken on the National Computational Infrastructure supercomputer located at the Australian National University. The research has made extensive use of the VisIt visualization and analysis tool VisIt (https://wci.llnl.gov/codes/visit). We thank Professor Gregory Taylor for providing us with the radio data of Hydra A used for Fig.~\ref{fig1} and the analysis in \S~\ref{s:b_r}.  We acknowledge the referee's constructive comments, which assisted the overall presentation of this paper.

\bsp

\label{lastpage}

\end{document}